\documentclass[11pt,a4paper]{article}
\pdfoutput=1
\usepackage{jheppub}
\usepackage{amsmath,amssymb}
\usepackage{graphicx}
\usepackage{dcolumn}
\usepackage{bm}
\usepackage{cancel}

\begin{document}

\title{New Higgs Production Mechanism \\ in Composite Higgs Models}

\date{\today}

\author{Adri\'an Carmona,}
\author{Mikael Chala}
\author{and Jos\'e Santiago}
\affiliation{CAFPE and Dpto. de F\'{\i}sica Te\'orica y del Cosmos \\
Universidad de Granada, E-18071, Granada, Spain}

\abstract{
Composite Higgs models are only now starting to be probed at the Large
Hadron Collider by Higgs searches. We point out that new resonances,
abundant in these models, can mediate new production
mechanisms for the composite Higgs. The new channels involve the
exchange of a massive color octet 
and single production of new fermion
resonances 
with subsequent decays into the Higgs and a Standard Model
quark. The sizable cross section and very distinctive kinematics allow
for a very clean extraction of the signal over the background with
high statistical significance. Heavy gluon masses up to $2.8$ 
TeV can be probed with data collected during 2012 and up to $5$ TeV
after the energy upgrade to $\sqrt{s}=14$ TeV. 
}

\maketitle

\section{Introduction}

Models of strong electroweak symmetry breaking (EWSB), 
and in particular composite
Higgs models~\cite{Kaplan:1983fs,Kaplan:1983sm,Dimopoulos:1981xc} 
in which the Higgs is a pseudo-Goldstone boson of a
global symmetry of the composite sector, are very attractive
candidates for a natural realization of EWSB.
Despite the expected abundance of new resonances at the TeV scale in
these models, because of the strong coupling in the composite sector,
electroweak precision tests require these new resonances to be
typically beyond the reach of current searches at the Large Hadron
Collider (LHC)~\cite{Agashe:2005dk,Carena:2006bn,Carena:2007ua}. The
Higgs couplings are 
modified in these models and thanks to the current ``Higgs
effort'' the LHC community is going through, the first non-trivial
LHC constraints on minimal composite Higgs models come from Higgs
searches, even for a moderate degree of
compositeness~\cite{Espinosa:2010vn,Espinosa:2012qj,
Carmi:2012yp,Azatov:2012bz,Espinosa:2012ir} (see
also~\cite{Contino:Trieste2012}).~\footnote{Genuine tests of Higgs
  compositeness, 
  based 
  on longitudinal gauge boson and Higgs scattering will
  require a much longer wait~\cite{Contino:2010mh}.}

New composite states, even if they are heavy enough to escape standard
analyses, could be eventually accessible with more ingenuous searches.
For instance, it has been recently emphasized that new color octet
resonances in composite models 
could be more efficiently searched for through their decay
into a massive fermion resonance and a Standard Model (SM)
quark~\cite{Vignaroli:2011ik,Barcelo:2011vk,Barcelo:2011wu,Bini:2011zb}
(this was first pointed out, but not fully explored in this context,
in~\cite{Dobrescu:2009vz}). 
These studies considered the channels in which the new fermion
resonance decays into an electroweak gauge boson and a SM quark.
However, a sizable fraction of the heavy fermion resonances will
decay into a SM quark and the composite Higgs, thus making this
process a new production mechanism for the composite Higgs.
We explore such production mechanism in this article
(see contribution 12 of~\cite{Brooijmans:2012yi} for preliminary results in the $Ht\bar{t}$
channel). The resulting
final state is of the form $H q\bar{q}$, with $q$ any SM quark. The
corresponding production cross-sections are sizable but not
dramatic. However, the very distinctive kinematics make the signal
cross section stand out of the background even with very simple (and
therefore quite robust) analyses. 
In the following we will call the new fermion resonances vector-like
quarks and the new color octet boson resonances heavy gluons. This
just reflects their nature and 
emphasizes that the new production mechanism occurs in more
general models than composite Higgs models. New color
octet resonances play no role in models of strong EWSB and are usually
disregarded. However, in the context of partial
compositeness~\cite{Kaplan:1991dc,
Contino:2006nn}, that we
adopt in this article, their presence is almost unavoidable.~\footnote{Similar
arguments have been used in~\cite{Falkowski:2008yr,Redi:2011zi}.}
Indeed, partial
compositeness implies a linear coupling of elementary fields to composite
operators of the strongly coupled sector. In particular, the linear coupling of
quarks implies that the strong sector must have a global $SU(3)_c$ symmetry.
The two point correlators of the corresponding conserved currents will include
vector resonances in the octet representation with which the SM gluons will
mix. Thus, although we are motivated by holographic models of strong EWSB,
the presence of new colored resonances is more general than that in the
context of models with partial compositeness.

Higgs production through the decay of new vector-like quarks has been
considered in the literature for quite some
time~\cite{delAguila:1989ba,delAguila:1989rq,AguilarSaavedra:2006gw,AguilarSaavedra:2009es,
Kribs:2010ii,Azatov:2012rj}.
In most
cases the process considered is pair production of the new vector-like
quarks followed by decays into electroweak and/or Higgs bosons and SM
quarks. This mechanism is essentially model independent as the
production is dominated by QCD and the equivalence theorem guarantees
that $\sim 1/4$ of the produced quarks decay into the Higgs and a SM
quark. The number of vector-like quarks being essentially the only
free parameter.~\footnote{Depending on the new quark quantum numbers,
  branching fractions into Higgs could be up to $1$ but only at the expense
  of having other quarks for which such channel is forbidden in such a
way that the global $1/4$ factor is approximately preserved.} An
alternative channel has been considered
in~\cite{Azuelos:2004dm,Vignaroli:2012sf}.
It consists of electroweak single
production of new vector-like quarks with subsequent decay into the
Higgs and a SM quark. This channel is more model dependent as the
production cross section depends on unknown electroweak couplings of
the heavy quark. All these processes are also present in the models we
are considering. The interesting fact is that all three are
sensitive to different couplings and it is therefore important to
study all of them independently as they can provide very useful
information on the properties of the composite sector.

The outline of the article is as follows. We introduce the model
and describe the main features relevant for the new Higgs production mechanism
in section~\ref{model}. Current constraints on the model are described
in section~\ref{constraints}. The analyses we propose to search for these
new channels are described in section~\ref{analyses} and we discuss
the results in section~\ref{results}. We summarize
our results in section~\ref{conclusions}.

\section{The Model\label{model}}

The new Higgs production mechanism we want to study consists of the
single production of a new vector-like quark (in association with a SM
quark) mediated by the exchange of a heavy gluon and with subsequent
decay in a SM quark and the Higgs. 
Thus, the only relevant ingredients are new vector-like
quarks and new massive gluons. This mechanism is therefore common to many
models of strong EWSB independently of whether the Higgs is a
pseudo-Goldstone boson or 
not. However, in order to be able to give quantitative results, we
focus in this article on the minimal composite Higgs model based on
the $SO(5)/SO(4)$ coset with composite fermions transforming
in the fundamental ($5$) representation of $SO(5)$, denoted by
MCHM$_5$~\cite{Agashe:2004rs,Contino:2006qr}. 
The coset structure and the fermion quantum numbers fix the Higgs
couplings to the SM particles (assuming the composite states to be heavy
enough so that mixing effects can be neglected) in terms of a single parameter
\begin{equation}
\xi=\frac{v^2}{f^2},
\end{equation}
where $v=2 m_W/g \approx 246$ GeV (with $m_W$ the $W$ mass and $g$ the
$SU(2)_L$ coupling constant) and $f$ is the decay constant of the
composite sector. In the MCHM$_5$ model, the ratios of the tree level
couplings of the Higgs to two SM particles to the corresponding SM
coupling read (see for
instance~\cite{Contino:2010mh,Espinosa:2010vn}): 
\begin{equation}
R_{HVV}\equiv \frac{g_{HVV}}{g^{\mathrm{SM}}_{HVV}}=\sqrt{1-\xi},\quad 
R_{Hff}\equiv \frac{g_{Hff}}{g^{\mathrm{SM}}_{Hff}}=\frac{1-2\xi}{\sqrt{1-\xi}},
\end{equation}
where $V$ and $f$ stand for any electroweak gauge boson and SM fermion,
respectively. The Higgs production cross
section receives a suppression proportional to these same factors:
\begin{equation}
\frac{\sigma(gg\to H)}
{\sigma(gg\to H)_{\textrm{SM}}}=R_{Hff}^2,\quad
\frac{\sigma(qq\to qq H)}
{\sigma(qq\to qqH)_{\textrm{SM}}}(\textrm{VBF})=
\frac{\sigma(qq\to V H)}
{\sigma(qq\to VH)_{\textrm{SM}}}=
R_{HVV}^2,
\end{equation}
where VBF stands for ``vector-boson fusion'' and we have included the
production processes relevant for the discussion in this article.
The different decay widths scale with the corresponding couplings squared
except for the $H\to \gamma \gamma$ channel that reads
\begin{equation}
\Gamma(H\to \gamma\gamma)
=\frac{(R_{Hff} I_\gamma + R_{HVV}
  J_\gamma)^2}
{(I_\gamma+J_\gamma)^2}
\Gamma^{\textrm{SM}}(H\to \gamma\gamma),
\end{equation}
where
\begin{equation}
I_\gamma=-\frac{8}{3} x_t[1+(1-x_t)f(x_t)],
\quad
J_\gamma=2+3x_W[1-(2-x_W)f(x_W)],
\end{equation}
with $x_t=4 m_t^2/m_H^2$, $x_W=4m_W^2/m_H^2$ and
\begin{equation}
f(x)=\left\{\begin{array}{ll}
\arcsin[1/\sqrt{x}]^2, & x\geq 1, \\
-\frac{1}{4}\left[\log
  \frac{1+\sqrt{1-x}}{1-\sqrt{1-x}}-\mathrm{i}\pi\right]^2, &
x<1.
\end{array}\right.
\end{equation}
The $H\to \gamma Z$ is also modified in a similar way, with different
loop functions. We do not give the explicit result as it will not be
used in the following.~\footnote{The $H\to \gamma \gamma$ has this simple
  form because of cancellations due to the pseudo-Goldstone nature of
  the Higgs and the fact that a unique flavor structure is present in
  the model. See~\cite{Azatov:2011qy} for a detailed discussion
  and~\cite{Casagrande:2010si,Azatov:2010pf,Carena:2012fk} for
  other cases.} 
The above equations completely determine all the relevant properties
of the Higgs due to its pseudo-Goldstone nature (parameterized by the
coefficient $\xi$). The Higgs becomes SM-like in the limit $\xi=0$.
\begin{figure}[ht]
\begin{center}
\includegraphics[width=0.6\textwidth,clip=]{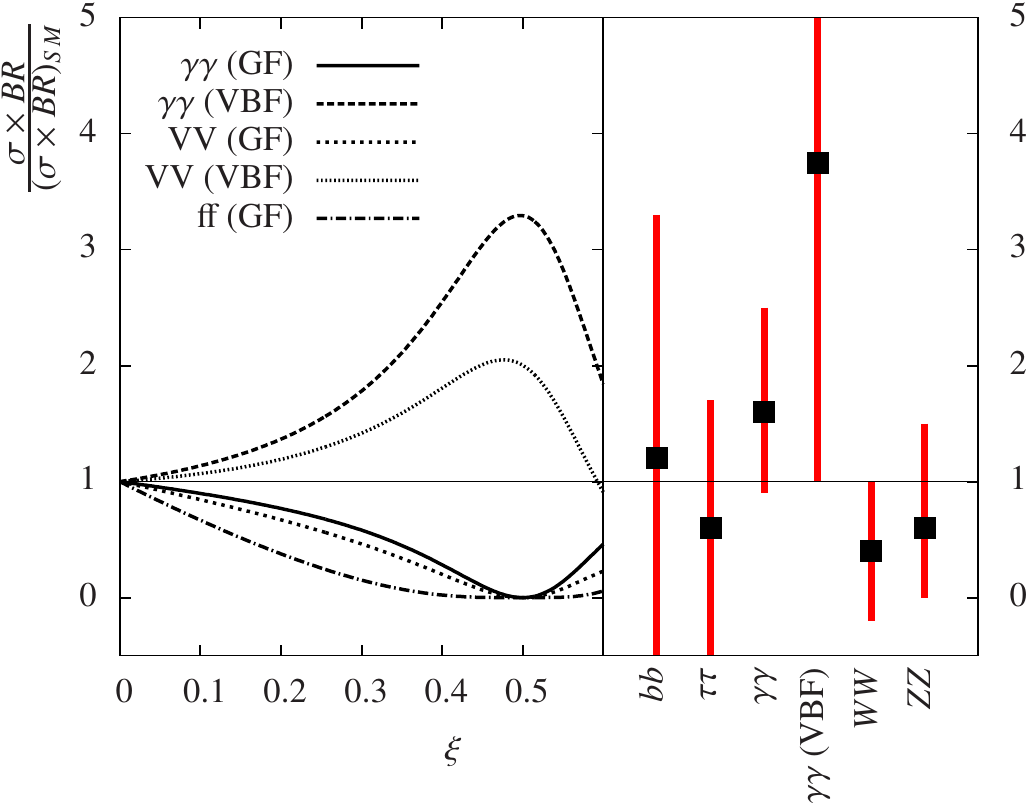} 
\caption{
Left panel: Higgs production cross section times branching ratio into
different channels in units of the corresponding SM process
(Higgs production has been separated in gluon fusion and
vector boson fusion, denoted by GF and VBF, respectively). 
Right panel: best fit value for the same observable obtained by the CMS
collaboration~\cite{CMS-PAS-HIG-12-001,CMS-PAS-HIG-12-008}. 
Both plots are for $m_H=125$ GeV.
}
\label{channels:fig}
\end{center}
\end{figure}
As an example, we show in Fig.~\ref{channels:fig} the Higgs
production cross section times branching ratio into different channels
in units of the corresponding SM cross section as a function of
$\xi$ for the case of a Higgs mass $m_H=125$ GeV. For comparison we
also show the best fit values for these cross sections as recently
reported by CMS.

Once we have discussed the effects of Higgs compositeness on standard 
Higgs searches we can turn to the ingredients present in our new
production mechanism.
Since most likely only the first
level of resonances will be accessible at the LHC, we use the two-site
version of the MCHM$_5$~\cite{Contino:2006nn,Bini:2011zb} to study its LHC
implications.~\footnote{Strictly speaking, we 
  use the deconstruction only for the collider search
  implications. As we have discussed, 
the features derived from the pseudo-Goldstone nature
  of the Higgs, which could be obtained in proper deconstructions of
  composite Higgs models~\cite{Panico:2011pw,DeCurtis:2011yx}, have
  been taken directly from the 
  holographic model in~\cite{Contino:2006qr}.} All the details of
the model can be found in the
original reference~\cite{Bini:2011zb}. Here we will just describe the
features that are directly relevant for the Higgs
production mechanism we want to study, namely the new massive gluons
and vector-like quarks present in the spectrum, together with their
couplings. The relevant new vector-like quarks are, for each family, 
two electroweak doublets
$Q_{1/6}$ and $Q_{7/6}$ of hypercharges $1/6$ and $7/6$, respectively 
and one singlet, $\tilde{T}$, of hypercharge $2/3$ 
\begin{equation}
Q_{1/6}^{(i)}=
\begin{pmatrix} 
T^{(i)} \\ B^{(i)} 
\end{pmatrix},\qquad
Q_{7/6}^{(i)}=
\begin{pmatrix} T^{(i)}_{5/3} \\  T^{(i)}_{2/3} \end{pmatrix}, \qquad
\tilde{T}^{(i)},\label{new:fermions}
\end{equation}
with masses $M_{Q_{1/6}^{(i)}}$,  $M_{Q_{7/6}^{(i)}}$,  $M_{T^{(i)}}$. 
There is also a massive gluon in the spectrum, a color octet
vector boson denoted by $G$, with mass $M_G$. 

The couplings of all the different
particles are fixed by the couplings in the composite sector and the
degree of compositeness of the different fields. The relevant
couplings in the composite sector are the Yukawa couplings of the up
and down sectors, that we take equal for simplicity, and the composite
coupling of the heavy gluons
\begin{equation}
Y_{\ast}\equiv Y_{\ast\,U}= Y_{\ast\,D},\quad g_{\ast\,3}.
\end{equation}
The degree of compositeness of each SM field can be parameterized by
a mixing angle, denoting the degree of compositeness of the SM gluon
($\theta_3$), left-handed (LH)
doublets ($\phi_q^{(i)}$), charge 
$2/3$ singlets ($\phi_{u}^{(i)}$) and charge $-1/3$ singlets
($\phi_{d}^{(i)}$). Finally, in the MCHM$_5$ model, there is an
extra parameter describing the way the LH doublet is split between two
sectors, denoted by the (small) angle $\phi_2^{(i)}$.
Of course, not all of these parameters are independent. For instance
the mass of the $Q_{1/6}$ is fixed in terms of the mass of $Q_{7/6}$
and the compositeness of the LH doublets
\begin{equation}
M_{Q_{1/6}^{(i)}}= M_{Q_{7/6}^{(i)}} /\cos \phi_q^{(i)}.
\end{equation}
Similarly we have
\begin{equation}
g_3= g_{\ast\,3}\sin \theta_3,
\end{equation}
with $g_3$ the SM QCD coupling, and
\begin{equation}
m_{u^{(i)}}\approx \frac{v}{\sqrt{2}} Y_{\ast U} \sin \phi_q^{(i)}
\sin \phi^{(i)}_{u}, \qquad
m_{d^{(i)}}\approx \frac{v}{\sqrt{2}} Y_{\ast D} \sin \phi_2^{(i)}
\sin \phi^{(i)}_{d}, \label{masas:SM}
\end{equation}
where $m_{u,d^{(i)}}$ are the corresponding SM quark masses.
Thus, we can use as independent parameters
\begin{equation}
g_{\ast\,3},~Y_\ast,~M_{Q_{7/6}^{(i)}},~M_{\tilde{T}^{(i)}},~\sin\phi_{u}^{(i)}\equiv
s_u^{(i)},\mbox{ and }\sin \phi_2^{(i)}\equiv s_2^{(i)},
\end{equation}
and compute all the other parameters in terms of these from the
equations above.

The coupling of the massive gluon
$G$ to the SM fermions is given by
\begin{equation}
g_{G\psi \psi}=g_3(\sin^2 \phi_\psi \cot \theta_3-\cos^2 \phi_\psi \tan \theta_3),
\end{equation}
where $\phi_\psi$ is one of the $\phi_q^{(i)}$, $\phi_{u}^{(i)}$ or
$\phi_{d}^{(i)}$, depending on the SM fermion involved. The
couplings of the heavy gluon to one SM fermion and one composite
resonance are given by
\begin{equation}
g_{G\psi \Psi}=g_3\frac{\sin\phi_\psi \cos \phi_\psi}{\sin \theta_3 \cos \theta_3},
\end{equation}
where the relevant combinations of $\psi \Psi$ are 
$u_L^{(i)} T_L^{(i)},~ d_L^{(i)} B_L^{(i)},~u_R^{(i)}
\tilde{T}_R^{(i)}$.\footnote{A charge $-1/3$ electroweak singlet can be also
produced in association with $d_R^{(i)}$ but this process is not
relevant for our Higgs production mechanism in the region of parameter
space we are interested in.} Finally, the coupling to two massive
resonances has the form
\begin{equation}
g_{G\Psi\Psi}=g_3 (\cos^2 \phi_\Psi \cot \theta_3 - \sin^2\phi_\Psi
\tan\theta_3), 
\end{equation}
where $\phi_\Psi=\phi_q,\phi_q,\phi_{u}$ for $Q=T,B,\tilde{T}$,
respectively.\footnote{Other quarks present in the spectrum couple in
pairs to $G$ with a similar structure but different values of the
couplings. These couplings are nevertheless irrelevant for the process
we are interested in, see Ref.~\cite{Bini:2011zb} for details.\label{other:Qs}}

If the composite sector is strongly coupled, $g_{\ast\,3}\gg 1$, we
have $\cot \theta_3\gg 1$ and the heavy resonances are strongly
coupled to the heavy gluon (except for maximally composite SM
fermions). This large coupling and the large multiplicity (as
mentioned in footnote~\ref{other:Qs}, there is a number of other
massive resonances with coupling $g_3 \cot \theta_3$) imply a very
large contribution to the heavy gluon width. 
For large values of the mass the width is of the order of
the mass itself and talking about resonances stops making sense. 
\begin{figure}[ht]
\begin{center}
\includegraphics[width=0.49\textwidth,clip=]{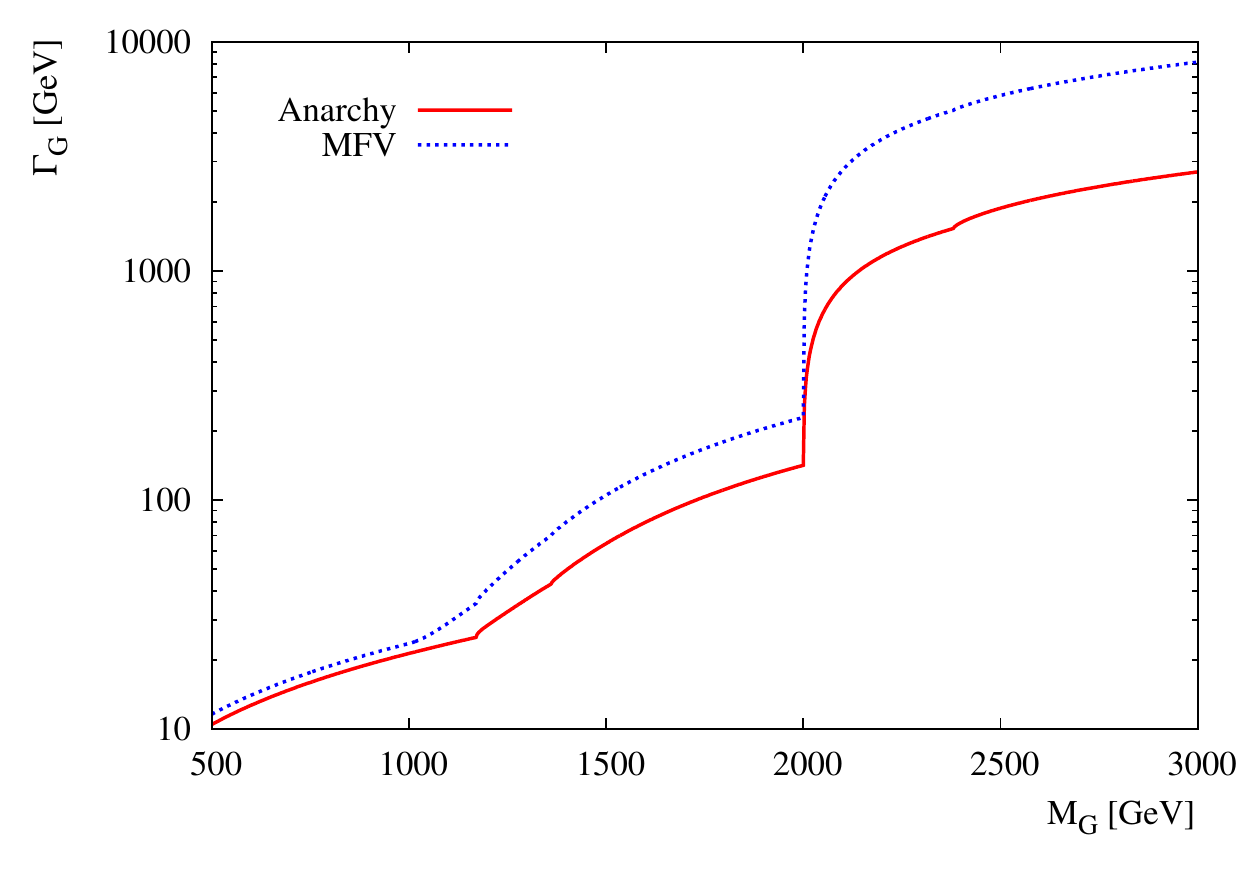} 
\hfil 
\includegraphics[width=0.49\textwidth,clip=]{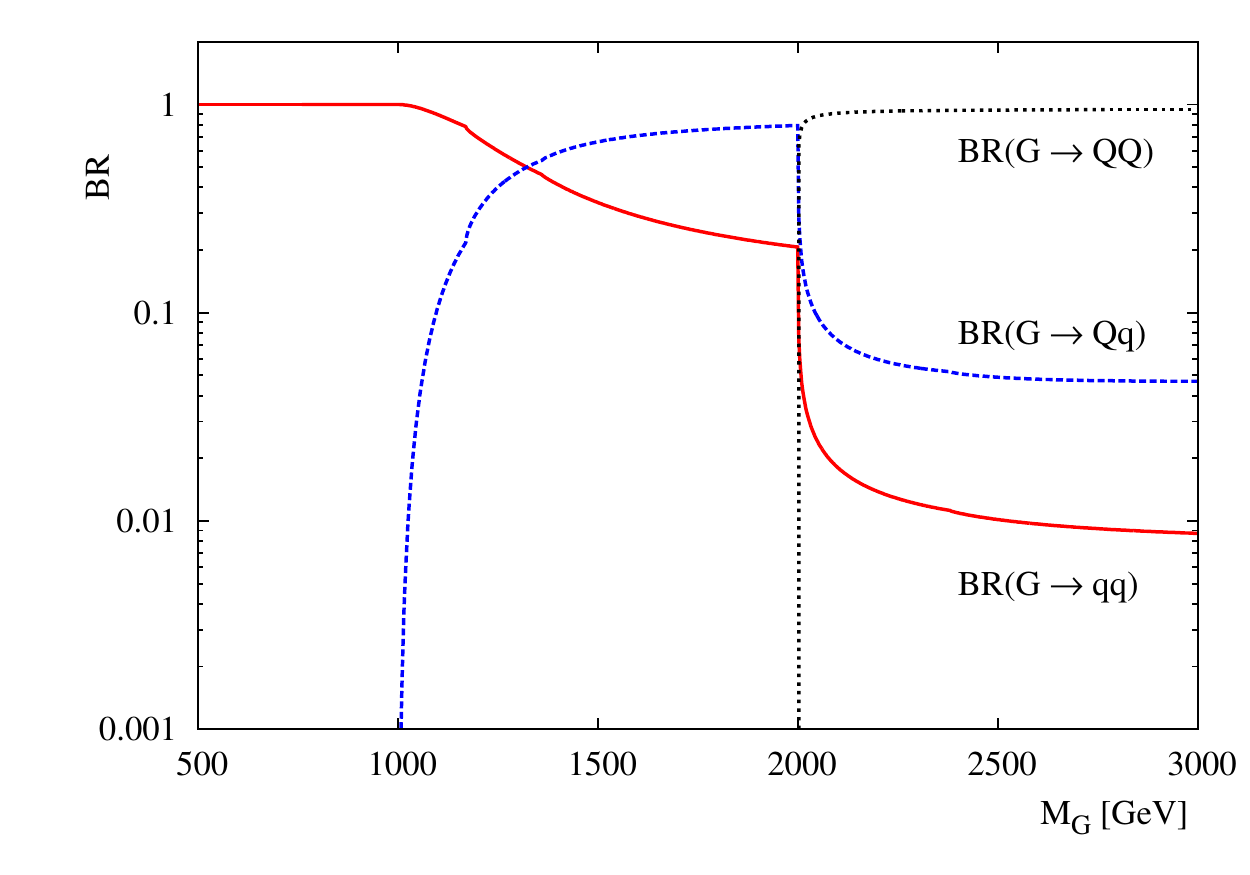} 
\caption{
Left panel: massive gluon width as a function of its mass for the
anarchic and MFV scenarios with $M_F=1$ TeV,
$s_u=0.6$, $s_2=0.1$ and
$g_{3\,\ast}=Y_\ast=3$ (in the anarchy scenario these values
correspond to third generation quarks only, see text for details). 
Right panel: massive gluon branching fraction in two SM quarks
(labeled $qq$), one SM
and one heavy quark ($Qq$) and two heavy quarks ($QQ$), respectively,
for the MFV scenario.
}
\label{widthandBR:fig}
\end{center}
\end{figure}
We
show in Fig.~\ref{widthandBR:fig} the massive gluon width (left panel)
and branching ratios (right panel) as a
function of its mass for $M_{Q_{7/6}}=M_{\tilde{T}}=1$ TeV, $s_u=0.6$,
$s_2=0.1$ and $g_{\ast\,3}=Y_\ast=3$. In the left panel we have chosen
two benchmark realizations of flavor. The first one corresponds to the
standard anarchy scenario
\begin{equation}
\mbox{Anarchy}\quad (s_{u}^{(1)}\ll s_{u}^{(2)}\ll
s_{u}^{(3)}\approx 1, \quad M^{(1,2)}\gg M^{(3)}),\label{anarchy:eq} 
\end{equation}
whereas the second one is the Minimal Flavor Violation (MFV)
of~\cite{Redi:2011zi} with mostly $u_R$ compositeness (see
also~\cite{Delaunay:2010dw,Delaunay:2011vv,Redi:2012uj} for
alternatives) 
\begin{equation}
\mbox{MFV}\quad  (s_{u}^{(1)}= s_{u}^{(2)}=
s_{u}^{(3)}\approx 1, \quad M^{(1,2)}= M^{(3)}).\label{MFV:eq} 
\end{equation}
where in both cases $M$ denote generically the fermion resonance
masses. 
(Strictly speaking, in the MFV scenario only the
$M_{\tilde{T}}$ masses have to be all equal. For simplicity we have
assumed all of them to be family independent.)
Due to the extreme widths developed when the decays into two massive
fermionic resonances open up, we restrict ourselves to the region of
parameter space in which these decay modes are kinematically
suppressed (see~\cite{Carena:2007tn} for an analysis of the case in
which these decays are allowed). Thus, in the following we fix
\begin{equation}
M_F\equiv M_{Q_{7/6}}=M_{\tilde{T}}=M_G/2.\label{fermion:masses}
\end{equation}
Also, although we will study the new Higgs production mechanism as a
function of the different input parameters, we will often report
results for a benchmark model, defined as
\begin{equation}
\mbox{Benchmark Model:}~M_F=M_G/2,~s_u=0.6,~g_{\ast\,3}=Y_\ast=3,~s_2=0.1,
\label{BM}
\end{equation}
where the different coefficients refer to just the third generation in
the anarchic case and to all three generations in the MFV scenario.
Naturalness arguments and the recent hints for a light Higgs might
prefer lighter fermion resonances for the third generation~\cite{Matsedonskyi:2012ym,Redi:2012ha,Marzocca:2012zn} (see also~\cite{Berger:2012ec}). In
these cases, if the heavy gluons are present, their width can easily
exceed the perturbative limit. For example if we fix $M_G=3$ TeV,
$s_u=0.6$, $s_2=0.1$ and $g_{\ast\, 3}=Y_\ast=3$ in the anarchic
scenario we get $\Gamma_G \gtrsim 0.9 M_G $ for $M_F\lesssim 1$ TeV.

\section{Experimental Constraints\label{constraints}}

Let us discuss current constraints on the model under consideration.
Direct searches of new states impose only mild constraints on the
parameter space allowed by electroweak precision tests. Nevertheless,
the increasing precision of the experimental searches can have some
impact on the parameter space as we describe here. There are four main
types of searches with implications in our model. The first one is
current Higgs searches, that are already starting to constrain the
parameter space of composite Higgs models. The second one is searches
for single production of new vector-like quarks that couple strongly
to first generation SM quarks. The last two involve
searches for new particles in $t\bar{t}$ and dijet final states,
respectively.  Processes leading to four-top final states represent a
complementary probe of these
models~\cite{Dicus:1994sw,Dobrescu:2007yp,Lillie:2007hd,Kumar:2009vs,Servant:2010zz,Perelstein:2011ez,Cacciapaglia:2011kz},
but they are not sensitive to the 
range of masses we are considering
here~\cite{AguilarSaavedra:2011ck,Zhou:2012dz}. 

\subsection{Higgs searches}

The implications of current Higgs searches on composite Higgs models
have been studied in detail in~\cite{Espinosa:2010vn,Espinosa:2012qj}
and in more general extensions
in~\cite{Carmi:2012yp,Azatov:2012bz,Espinosa:2012ir,Giardino:2012ww} (see
also~\cite{Contino:Trieste2012}). The result is that, for $m_H=125$ GeV, the region
\begin{equation}
0\leq \xi \lesssim 0.4,
\end{equation}
is allowed. Other masses are also allowed for certain values of the
degree of compositeness (for instance $m_H\approx 130$ GeV is allowed
for $\xi \gtrsim 0.2-0.3$). We will use as benchmark values $m_H=125$
GeV and $\xi=0.2$ in this article but will also consider the effect of
variations in these parameters.

\subsection{Vector-like quark searches}

Our assumption on the masses of the fermion resonances,
Eq. (\ref{fermion:masses}), was imposed to avoid too large a width for
the massive gluon. Electroweak precision tests already impose stringent
bounds on the masses of the heavy gluons (assuming that their mass is
similar to the one of electroweak vector resonances), making the fermion
resonances heavier than the reach of current searches. The only
exception occurs in the case of fermion resonances that mix strongly
with first generation quarks, as happens in the MFV scenario we are
considering. It was noted in Refs.~\cite{Atre:2008iu,Atre:2011ae} (see
also \cite{delAguila:2010es} for lepton resonances) that dedicated
searches at hadron colliders of electroweak single production of these
fermion resonances could probe quite large masses. The main reason is
the large cross section due to the presence of valence quarks in the
initial state and the distinctive kinematics. In fact, ATLAS has
recently performed such a search on their data~\cite{Aad:2011yn} 
and as we discuss now, their
null results imply the most stringent constraints for the MFV scenario
in a large fraction of parameter space.

Ref.~\cite{Aad:2011yn} reports
$95\%$ CL limits on the coupling $\tilde{\kappa}_{uU,uD}$ defined from
the general parameterization of the gauge couplings
\begin{equation}
\frac{g}{2c_W}\tilde{\kappa}_{uU} \frac{v}{m_U} Z_\mu \bar{u}_R
\gamma^\mu U_R + \frac{g}{\sqrt{2}} \tilde{\kappa}_{uD}
\frac{v}{m_D} W^+_\mu \bar{u}_R \gamma^\mu D_R + \mathrm{h.c.},
\end{equation}
where $g$ is the $SU(2)_L$ coupling, $c_W$ the cosine of the weak
angle and $m_{U,D}$ are the masses of new vector-like quarks, $U$
and $D$, of charges $2/3$ and $-1/3$, respectively. The bounds on
$\tilde{\kappa}^2$ are reported assuming only one type of quark at
each time. This applies directly to our model in the case of the
neutral current channel ($\tilde{\kappa}_{uU}$). In the charged
current channel, however, we have two new vector-like quarks,
$B^{(1)}$ and $T^{(1)}_{5/3}$ in the notation of Eq. (\ref{new:fermions}),
instead of just one and the charge $5/3$ one has a production
cross-section (which is proportional to $\tilde{\kappa}^2$)
approximately two times the one of the charge $-1/3$
quark~\cite{Atre:2011ae}. The net
result is that the bounds from Ref.~\cite{Aad:2011yn} on
$\tilde{\kappa}_{uD}^2$ are in practice a factor of 3 stronger when
interpreted in our model.

\begin{figure}[ht]
\begin{center}
\includegraphics[width=0.6\textwidth,clip=]{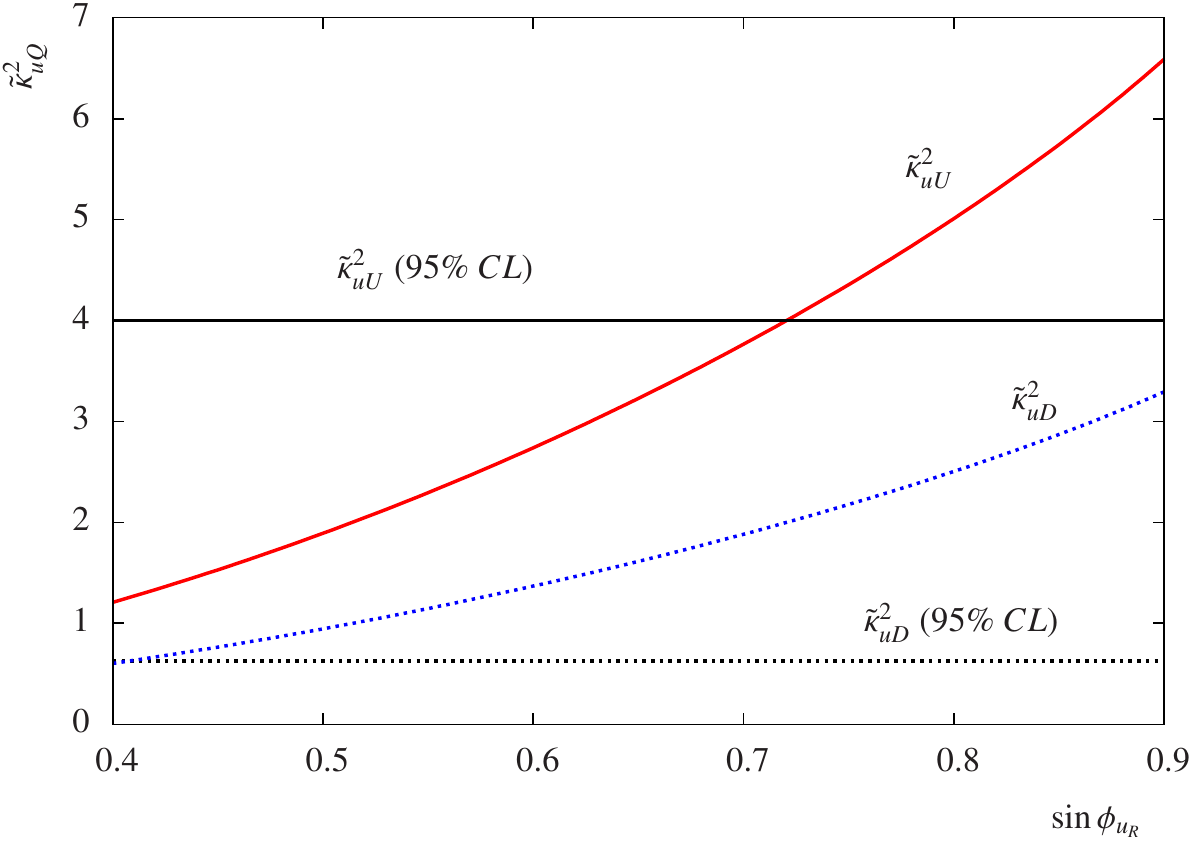} 
\caption{Values of $\tilde{\kappa}_{uU,uD}^2$ obtained in the MFV
  scenario with $Y_{\ast\,U}=3$ and $M_F=1$. The horizontal lines
  correspond to the
  $95\%$ CL experimental upper bounds as interpreted in our model.
}
\label{kappa:fig}
\end{center}
\end{figure}

The value of $\tilde{\kappa}_{uU,uD}$ can be easily computed in our
model, following the method outlined in section 13
of~\cite{Brooijmans:2008se}. It is
mostly sensitive to $s_{u}^{(1)}$ and $Y_{\ast\,U}$ although
the dependence on the latter is milder. As an example, we show the
value obtained in the MFV scenario with $Y_{\ast\, U}=3$ and $M_F=1$
TeV, as a function of $s_{u}^{(1)}$ in Fig.~\ref{kappa:fig}, together
with the corresponding experimental bounds as interpreted in our
model. We see that for these values of the input parameters, only a
relatively small degree of compositeness $s_u^{(1)}\lesssim 0.4$
is allowed. Due to limited statistics, only fermion masses up to 1
TeV, could be constrained in~\cite{Aad:2011yn}. Thus, anything above that
is currently experimentally allowed but we see from the figure that
updates on these searches are likely to be one of the strongest
constraints on the degree of compositeness of the $u_R$ quark in the
MFV scenario.

\subsection{$t\bar{t}$ resonance searches}

In models of strong EWSB with a composite top, new vector resonances
and in particular heavy gluons decay most of the time in
$t\bar{t}$ pairs, which has been traditionally considered the golden
discovery mode of such new particles. If the top is not fully
composite, if other quarks can be as composite as the top (as happens
in the MFV scenario), or if new decay channels involving fermion composite
states are open, the branching fraction into $t\bar{t}$ can substantially
change.
As an example for the benchmark model in Eq.(\ref{BM}) with MFV
we have $BR(G\to t\bar{t})=0.036$ and the
$t\bar{t}$ production is at least an order of magnitude below current
limits~\cite{CMS-PAS-TOP-11-009,ATLAS-CONF-2011-123,ATLAS-CONF-2012-029}. We
will include the constraints resulting from these searches in our
general analysis in section~\ref{results}.

\subsection{Di-jet searches}

We are considering models that depart
from the standard composite models in two main aspects. The first is
that we do not necessarily consider a completely composite
$t_R$~\cite{Pomarol:2008bh}. The
second is that we also consider the MFV scenario in which the $u_R$
and $c_R$ are as composite as the $t_R$. In this case, dijet
production can impose stringent constraints in the
model~\cite{Lillie:2007ve,Redi:2011zi}. 
A very detailed study of the implication of
dijet searches on contact interactions has been recently reported
in~\cite{Domenech:2012ai}. Their original analysis, which considers the
early LHC data of Ref.~\cite{Aad:2011aj} with an integrated
luminosity of just 36 pb$^{-1}$, has been updated to include the
latest experimental results~\cite{Chatrchyan:2012bf} with an integrated luminosity of
2.2 fb$^{-1}$.~\footnote{We would like to thank O. Domenech,
  A. Pomarol and J. Serra for providing us with the updated analysis and for useful
discussions.} 
Denoting the coupling of the first generation SM quarks to the massive
gluon by
\begin{equation}
G_\mu^A \Big[g_{q_L} \bar{q}_L \gamma^\mu T^A q_L
+g_{u_R} \bar{u}_R \gamma^\mu T^A u_R
+g_{d_R} \bar{d}_R \gamma^\mu T^A d_R],
\end{equation}
we get the following effective Lagrangian after integration of the massive
gluon, in the basis of~\cite{Domenech:2012ai}
\begin{equation}
\mathcal{L}=
\frac{c_{uu}^{(1)}}{M^2} \mathcal{O}_{uu}^{(1)}
+\frac{c_{dd}^{(1)}}{M^2} \mathcal{O}_{dd}^{(1)}
+\frac{c_{ud}^{(8)}}{M^2} \mathcal{O}_{ud}^{(8)}
+\frac{c_{qq}^{(8)}}{M^2} \mathcal{O}_{qq}^{(8)}
+\frac{c_{qu}^{(8)}}{M^2} \mathcal{O}_{qu}^{(8)}
+\frac{c_{qd}^{(8)}}{M^2} \mathcal{O}_{qd}^{(8)},
\end{equation}
where the different coefficients read
\begin{eqnarray}
c_{uu}^{(1)}&=&-\frac{g_{u_R}^2}{6}, \quad
c_{dd}^{(1)}=-\frac{g_{d_R}^2}{6}, 
\quad c_{ud}^{(8)}=-g_{u_R} g_{d_R}, \\
c_{qq}^{(8)}&=&-\frac{g_{q_L}^2}{2}, \quad
c_{qu}^{(8)}=-g_{q_L}g_{u_R}, \quad
c_{qd}^{(8)}=-g_{q_L} g_{d_R}.
\end{eqnarray}
The results of~\cite{Domenech:2012ai} can be directly applied to these coefficients
to obtain the corresponding bound on $M_G$.

Direct dijet resonance searches can also constrain our model. We have
simulated dijet signals in our model and compared the results after
cuts with the bounds on simplified gaussian resonances reported
in~\cite{Aad:2011fq}. The corresponding limits are included in our
final results.

\section{Searches for New Composite Higgs Production
  Mechanisms\label{analyses}}

\begin{figure}[ht]
\begin{center}
\includegraphics[height=3.3cm,clip=]{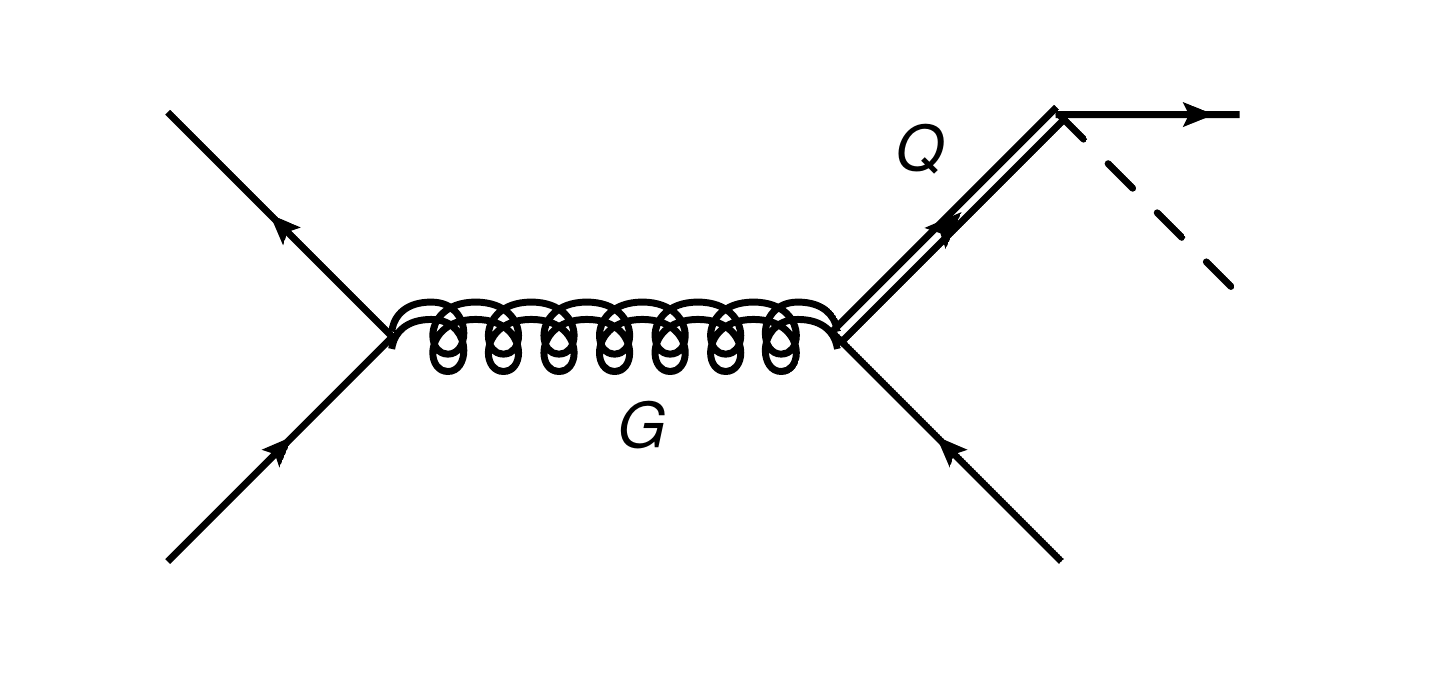} 
\hfil 
\includegraphics[height=2.5cm,clip=]{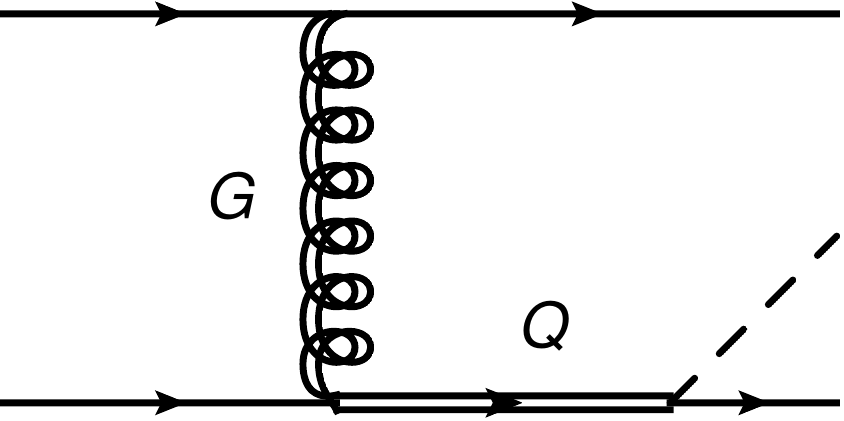} 
\caption{
Sample diagrams for the new Higgs production mechanism. The t-channel
exchange on the right is relevant for composite $u$ or $d$ quarks.
}
\label{production:diagram}
\end{center}
\end{figure}

\begin{figure}[ht]
\begin{center}
\includegraphics[width=0.45\textwidth,clip=]{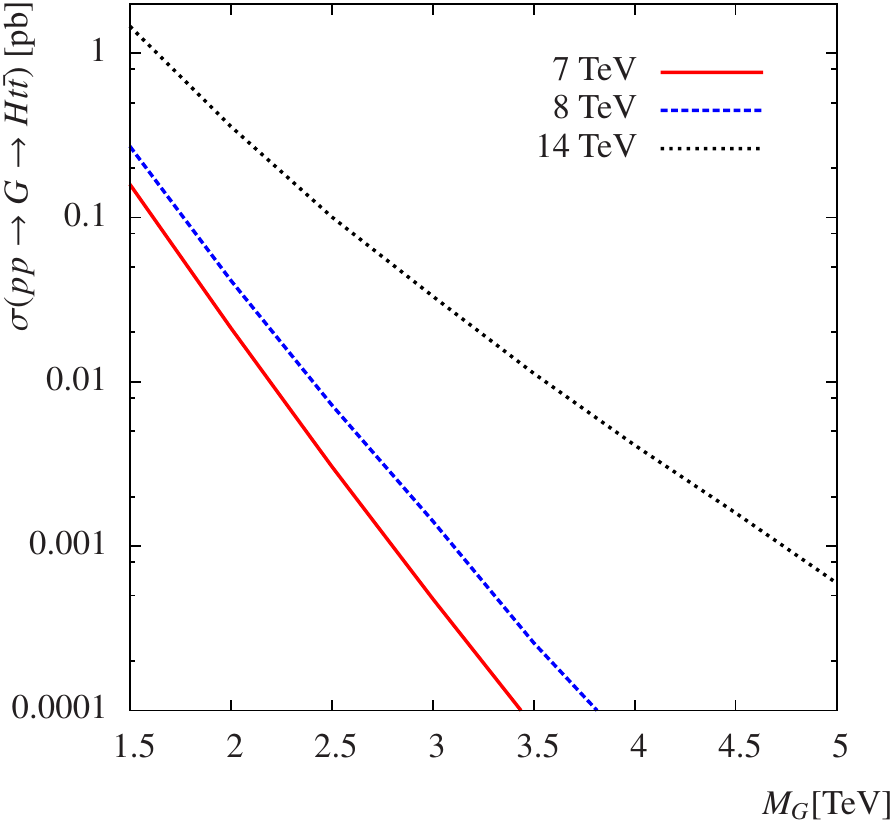} 
\hfil 
\includegraphics[width=0.45\textwidth,clip=]{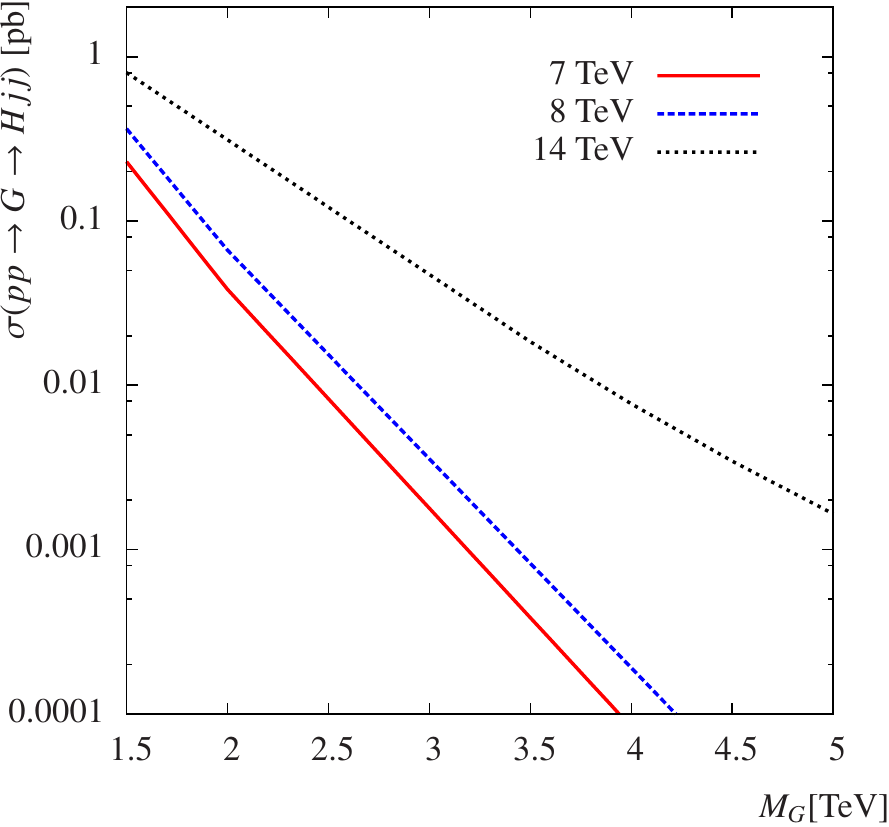} 
\caption{
Left panel: $Ht\bar{t}$ production cross section in the benchmark MFV
composite model mediated by a color octet vector resonance with decay
into a fermionic resonance and a top quark. Right panel: same for
the $Hjj$ channel. 
}
\label{qqHXsec:fig}
\end{center}
\end{figure}

The new Higgs production mechanism we want to study consists of single
production of new vector-like quarks (together with a SM quark)
mediated by the exchange of a heavy color octet vector boson. The
vector-like quark then decays into the composite Higgs and a SM
quark. Sample diagrams of this production mechanism are shown in
Fig.~\ref{production:diagram}. The t-channel diagram on the right
panel of the figure is only relevant for the MFV scenario in which
first generation SM quarks are strongly composite. 
The final state is either $Ht\bar{t}$ or $Hjj$ (the latter gives a
substantial contribution only in the MFV scenario). The corresponding
cross sections depend on the coupling of $G$ to the SM quarks, to $qQ$
and also on the branching fraction of the heavy quarks into a SM quark
and the Higgs. The
relevant such branching fractions are, in the limit of large masses~\cite{Bini:2011zb}
\begin{equation}
\mathrm{BR}(T^{(i)}\to u^{(i)} H)\approx 0.5,\quad
\mathrm{BR}(\tilde{T}^{(i)}\to u^{(i)} H)\approx 0.25.
\end{equation}
Other channels either do not result in a Higgs or their production is
strongly suppressed due to small degree of compositeness.
We show in Fig.~\ref{qqHXsec:fig} the production cross section times
branching ratio for the $Ht\bar{t}$ (left) and $Hjj$ (right) channels
in the benchmark model of Eq.(\ref{BM}) within the
MFV scenario. In the anarchic scenario the $Hjj$ channel is negligible
and the $Ht\bar{t}$ is enhanced by a factor $\approx 25-40 \%$,
depending on the value of $M_G$.

In the following we describe dedicated analyses for the $H t\bar{t}$
and $Hjj$ channels. We have considered three different configurations
for the LHC parameters, namely 5 fb$^{-1}$ integrated luminosity at
$\sqrt{s}=7$ TeV, 20 fb$^{-1}$ integrated luminosity at
$\sqrt{s}=8$ TeV and 100 fb$^{-1}$ integrated luminosity at
$\sqrt{s}=14$ TeV. The range of masses probed with the first two
configurations (that we call the low energy phase) is quite different
from the one probed by the high energy phase (the third option). Thus,
the analyses are also different depending on the phase. In particular,
as we describe below, the analysis of the $H t\bar{t}$ channel in the
high energy phase benefits from using boosted techniques. 

We have used \texttt{MADGRAPH V4.5.0}~\cite{Alwall:2007st} and \texttt{ALPGEN
  V.2.13}~\cite{Mangano:2002ea} to generate signal and background partonic
events. Such events have then been passed through \texttt{PYTHIA
  6.4}~\cite{Sjostrand:2006za} for hadronization and showering and \texttt{DELPHES
  V1.9}~\cite{Ovyn:2009tx} for detector simulation. Regarding the latter we have
used a tuned version of the standard ATLAS card that results in an
very good agreement with published experimental results. We have used
the CTEQ6L1 PDFs and the default values of the renormalization
and factorization scales. The
backgrounds have been matched using the MLM method. In our analyses we
define jets with a cone size $\Delta R=0.7$, $p_T(j)>30$ GeV and
$|\eta_j|<5$. Isolated charged leptons ($e$ or $\mu$) are considered
when $p_T(l)>20$ GeV and $|\eta_l|<2.5$. We have assumed a b-tagging
efficiency of 0.7 in our analyses.
Finally, we use as discriminating variable
\begin{equation}
S_T\equiv \sum_{j=1}^{n_j} p_T(j) + \sum_{l=1}^{n_l} p_T(l)+ \cancel{E}_T,
\end{equation}
where $n_{j,l}$ is the relevant number of jets or leptons
(ordered according to their $p_T$), which depends on the analysis and
will be specified below.
We list in Table~\ref{backgrounds:table} 
the main backgrounds and two sample points in parameter space for our
benchmark model together with their cross sections.
\begin{table}[ht]
\begin{center}
\begin{tabular}{|l|c|c|c|}
\hline
Process & LHC7  &  LHC8  &  LHC14 
 \\
 & $\sigma$ [pb]   &  $\sigma$ [pb]   &  $\sigma$ [pb]  
 \\
\hline
$Ht\bar{t}$ ($M_G=2$ TeV, MFV) & 0.0213 & 0.0414 & 0.270 \\ 
$Ht\bar{t}$ ($M_G=3$ TeV, MFV) & 0.00053& 0.0015 & 0.033 \\
$Ht\bar{t}$ ($M_G=2$ TeV, Anarchy) &0.029 & 0.058 & 0.39 \\  
$Ht\bar{t}$ ($M_G=3$ TeV, Anarchy) &0.00061 & 0.0018 & 0.046 \\ 
$Hjj$ ($M_G=2$ TeV, MFV) & 0.04 & 0.07& 0.44 \\ 
\hline
$t\bar{t}$+0-4 jets (semileptonic+leptonic) & 47.9 & 70.47 & 268.55 \\
$t\bar{t}b\bar{b}$  & 0.09  & 0.15 & 0.85 \\
$Z$+1-4 jets (leptonic) & 530.5 & 641 & 1423 \\
$WW$ + 0-2 jets (semileptonic+leptonic) & 15 & 22.6 & 49 \\
$W$+1-2 jets ($p_T> 150$ GeV, leptonic) & $-$ & $-$  & 84.9  \\
$W$+1-4 jets (leptonic)  & 5133 & 6489  & $-$  \\
\hline
\end{tabular}
\caption{Cross section for different the signal and main backgrounds
  for different values of the LHC energy. In $t\bar{t}$ and
  $t\bar{t}b\bar{b}$ leptonic decays refer to $e$ or $\mu$, in the
  other cases decays into $\tau$ are also included. The corresponding
  branching fractions are included in the calculation of the cross
  section (when the decays -leptonic or semileptonic- are explicitly
  stated). \label{backgrounds:table}
}
\end{center}
\end{table}

Given a number of signal ($s$) and background ($b$) events after the
corresponding cuts, we compute the statistical significance of the
signal from
\begin{equation}
\mathcal{S}(s,b)=\sqrt{2\times \left[ (s+b) \ln \left(
    1+\frac{s}{b}\right)
-s \right]}.\label{cl}
\end{equation} 
We describe in this section our suggested analyses and the effect they
have on signal and background for a specific point in parameter
space. We will then present our results in the next section as a
function of the input parameters.

\subsection{$H t\bar{t}$ analysis: low energy phase}

The mass range that can be probed at the LHC within the low energy
phase ($\sqrt{s}=7$ or $8$ TeV) is relatively low. This means that
the decay products are not extremelly boosted. We have found that
traditional analyses are more efficient probing this region of
parameter space than analysis that use boosted 
techniques. Also, since we have the leptonic top decays to trigger on,
we can afford to use the main Higgs decay channel, namely $b\bar{b}$,
with a branching ratio $\mathrm{BR}(H\to b\bar{b})=0.48$ for the
benchmark model in Eq.~(\ref{BM}) (the changes in the BR for different
values of the input parameters can be recovered using the equations in
Section~\ref{model} and are shown, for reference, in Fig.~\ref{HiggsBRvsxi:fig}). 
We are therefore interested in the following
process
\begin{equation}
pp\to G\to T \bar{t}+\bar{T}t \to H t \bar{t}\to 4b+2j
+l+\cancel{E}_T.
\label{Htt:process} 
\end{equation}
The main backgrounds are $t\bar{t}$ and $t\bar{t}b\bar{b}$. 
In order to reduce the number of
background events to manageable values we impose the following initial
cuts
\begin{itemize}
\item At least 4 jets, of which at least 3 must be tagged as b-jets.
\item At least 1 isolated charged lepton.
\item A cut on $S_T$ (in this case we have $n_j=4$ and $n_l=1$) 
that depends on the test $M_G$ we are considering
\begin{equation}
S_T>0.9,~1.1,~1.5\mbox{ TeV for }M_G=1.5,~2,~2.5\mbox{ TeV}.
\end{equation}
\end{itemize} 
\begin{table}[ht]
\begin{center}
\begin{tabular}{|c|c|c|c|}
\hline
\multicolumn{4}{|c|}{sample $Ht\bar{t}$ analysis (low energy phase)} \\
\hline
cut &  $\epsilon_{M_G=2~\mathrm{TeV}}$ & $\epsilon_{t\bar{t}}$ &
  $\epsilon_{t\bar{t}b\bar{b}}$ 
\\
\hline
$n_j \geq 4$ & 77.31 & 52.16  & 91.85 
\\ \hline
$n_l\geq 1$ & 66.86 &  63.02 &  42.84
\\ \hline
$n_b\geq 3$ & 35.31 &  2.64  & 33.08
\\ \hline
$S_T$ &       75.01 &  0.12 & 1.20
\\ \hline
Total &     13.69 & 0.00108& 0.156
\\ \hline
\end{tabular}
\hfil
\begin{tabular}{|c|c|c|c|}
\hline
\multicolumn{4}{|c|}{$Ht\bar{t}$ (low energy phase)} \\
\hline
$M_G$ [TeV] &  $\epsilon_s$ & $\epsilon_{t\bar{t}}$ &
  $\epsilon_{t\bar{t}b\bar{b}}$ 
\\
\hline
1.5 & 15.8 & 0.00652 & 0.514 
\\
\hline
2.0 & 13.69 & 0.00108 & 0.156
\\ \hline
2.5 & 9.67 & 0.000292 & 0.0174
\\ \hline
3.0 & 9.14 & 0.000292 & 0.0174
\\ \hline
\multicolumn{4}{c}{\phantom{Total}}
\end{tabular}
\caption{Left panel: Cut by cut efficiencies for the signal and main
  backgrounds for the $Ht\bar{t}$ analysis in the low energy phase for
  a sample point (benchmark model with MFV and $M_G=2$ TeV). Right
  panel: global efficiencies for the signal and relevant backgrounds
  as a function of $M_G$. All efficiencies are reported as per cent.}\label{efficiencies:Htt:lowE}
\end{center}
\end{table}
We show in the left panel of Table~\ref{efficiencies:Htt:lowE}, the efficiencies
of the 
different cuts for the main backgrounds and our signal for the MFV
realization of our benchmark model, Eq.~(\ref{BM}), with $M_G=2$
TeV. The global efficiencies for the signal (again BM with MFV) and
relevant backgrounds are reported, as a function of $M_G$ in the right
panel of the table.
From the numbers in this table and the cross sections for the
signal (times the $BR(H\to b\bar{b})=0.48$ for $m_H=125$ GeV and
$\xi=0.2$) 
and background reported in Table~\ref{backgrounds:table} we obtain 
a statistical significance of
\begin{eqnarray}
&&\mathcal{S}(7.1,3.3)=3.1\quad 
(\mathcal{L}=5~\mathrm{fb}^{-1},~\sqrt{s}=7\mbox{
  TeV},~M_G=2\mbox{ TeV, MFV}),
\\
&&\mathcal{S}(55.9,19.9)=10.8\quad
(\mathcal{L}=20~\mathrm{fb}^{-1},~\sqrt{s}=8\mbox{ TeV}
,~M_G=2\mbox{ TeV, MFV}).
\end{eqnarray}
In the case of $\sqrt{s}=8$ TeV a luminosity of just 4.3 fb$^{-1}$
would suffice for a $5~\sigma$ discovery.
In the anarchy scenario, there is a $\approx 35\%$ ($38\%$)
enhancement for $\sqrt{s}=7$ (8) TeV of the
signal, resulting in the following statistical significances
\begin{eqnarray}
&&\mathcal{S}(9.5,3.3)=4.\quad 
(\mathcal{L}=5~\mathrm{fb}^{-1},~\sqrt{s}=7\mbox{
  TeV},~M_G=2\mbox{ TeV, Anarchy}),
\\
&&\mathcal{S}(77.2,19.9)=13.9\quad
(\mathcal{L}=20~\mathrm{fb}^{-1},~\sqrt{s}=8\mbox{ TeV}
,~M_G=2\mbox{ TeV, Anarchy}).
\end{eqnarray}
As we will discuss in the next section, in which we describe our results as a
function of the input parameters, current constraints from
dijet contact interactions imply a bound $M_G\geq 2.5$ TeV for the
benchmark model in the anarchy scenario. 
These values cannot be probed with the 7 TeV run but
with $\sqrt{s}=8$ TeV it should be possible to discover (exclude) it with
20 (5) fb$^{-1}$.

\subsection{$H t\bar{t}$ analysis: high energy phase}

In the high energy phase, $\sqrt{s}=14$ TeV, larger masses can be
probed. In this case the decay products of $G$ and $Q$ are highly
boosted and one can benefit from the use of boosted techniques. In
this study we use a very simple technique, based on fat jet invariant
masses~\cite{Skiba:2007fw,Holdom:2007nw,Holdom:2007ap}. 
Clearly there is room for improvement if more
sophisticated tools are used~\cite{Abdesselam:2010pt,Altheimer:2012mn}.
The new set of cuts optimized for the larger masses probed are the
following
\begin{itemize}
\item At least 3 jets, with a minimum of 2 b tags.
\item At least 1 isolated charged lepton.
\item All jets are then ordered according to their invariant mass and
  the first two jets are required to have invariant masses close to
  the top and Higgs mass, respectively, $|m_{j_1}-m_t|\leq 40$ GeV and
  $|m_{j_2}-m_H|\leq 40$ GeV (here $j_{1,2}$ are the jets with the
  largest and second largest invariant masses).
\item A cut on $S_T$ (in this case we have $n_j=3$ and $n_l=1$) 
that depends on the test $M_G$ we are considering
\begin{equation}
S_T>1.2,~1.5,~1.7,~2\mbox{ TeV for }M_G=2,~2.5,~3,\geq 3.5\mbox{ TeV}.
\end{equation}
\end{itemize}
\begin{table}[ht]
\begin{center}
\begin{tabular}{|c|c|c|c|}
\hline
\multicolumn{4}{|c|}{sample $Ht\bar{t}$ analysis (high energy phase)} \\
\hline
cut &  $\epsilon_{M_G=3~\mathrm{TeV}}$ & $\epsilon_{t\bar{t}}$ &
  $\epsilon_{t\bar{t}b\bar{b}}$ 
\\
\hline
$n_j \geq 3$ & 98.03 & 85.46  & 98.88 
\\ \hline
$n_l\geq 1$ & 75.24 &  61.08 &  45.16
\\ \hline
$n_b\geq 2$ & 64.38 &  29.49  & 68.50
\\ \hline
$m_{j_1} \sim m_t$ &       58.08 &  0.22 & 1.70
\\ \hline
$m_{j_2}\sim m_h$ &       72.70 &  15.36 & 31.72
\\ \hline
$S_T$ &       90.07 &  10.24 & 18.10
\\ \hline
Total &     18.06 & 0.00054& 0.0298
\\ \hline
\end{tabular}
\hfil
\begin{tabular}{|c|c|c|c|}
\hline
\multicolumn{4}{|c|}{$Ht\bar{t}$ (high energy phase)} \\
\hline
$M_G$ [TeV] &  $\epsilon_s$ & $\epsilon_{t\bar{t}}$ &
  $\epsilon_{t\bar{t}b\bar{b}}$ 
\\
\hline
2.0 & 11.74 & 0.00265 & 0.1021
\\ \hline
2.5 & 15.61 & 0.00095 & 0.0518
\\ \hline
3.0 & 18.06 & 0.00054 & 0.0298
\\ \hline
3.5 & 17.74 & 0.00027 & 0.0188
\\ \hline
4 & 19.08 & 0.00027 & 0.0188
\\ \hline
4.5 & 19.40 & 0.00027 & 0.0188
\\ \hline
\multicolumn{4}{c}{\phantom{Total}}
\end{tabular}
\caption{Left panel: Cut by cut efficiencies for the signal and main
  backgrounds for the $Ht\bar{t}$ analysis in the high energy phase for
  a sample point (benchmark model with MFV and $M_G=3$ TeV). Right
  panel: global efficiencies for the signal and relevant backgrounds
  as a function of $M_G$. All efficiencies are reported as per cent.}\label{efficiencies:Htt:highE}
\end{center}
\end{table}
The results of these cuts on the main backgrounds and the signal are
reported in table~\ref{efficiencies:Htt:highE}. In the left panel we
report cut-by-cut efficiencies for a sample signal point (MFV
for the benchmark model with $M_G=3$ TeV) whereas in the right panel
we report the global efficiencies as a function of $M_G$.
The corresponding statistical
significance is
\begin{equation}
\mathcal{S}(288,170)=18\quad
(\mathcal{L}=100~\mathrm{fb}^{-1},~\sqrt{s}=14~\mathrm{TeV},~M_G=3\mbox{
  TeV, MFV}).
\end{equation}
A $5\sigma$ discovery could be reached with this energy using just an
integrated luminosify of 7.5 fb$^{-1}$.
The $38\%$ enhancement in the anarchy case results in
\begin{equation}
\mathcal{S}(398,170)=24\quad
(\mathcal{L}=100~\mathrm{fb}^{-1},~\sqrt{s}=14\mbox{ TeV}
,~M_G=3\mbox{ TeV, Anarchy}).
\end{equation}
Results for other
regions of parameter space will be reported in the next section.

\subsection{$H jj$ analysis: high energy phase}

The last analysis we are going to describe is relevant for the MFV
realization of flavor in which the SM quarks produced in association
with the Higgs are first and second generation quarks. The Higgs is then
produced in association with two jets with a cross section given, for
our benchmark model, in Fig.~\ref{channels:fig} (right panel). Even 
if the two extra jets are quite hard or forward, depending on whether
we have an $s-$ or $t-$channel contribution, the signal is completely
swamped by backgrounds if we consider the $H\to b\bar{b}$ decay
channel. We are therefore forced to consider the $H\to WW^\ast$
channel, with a $BR(H\to WW^\ast)=0.33$ for our benchmark model. Even
so, the relatively small cross sections and the huge $W+j$ background
makes the dilepton mode the only one in which the signal can
realistically extracted from the background. The penalty to pay is
then the low cross sections and we will be in all cases statistics limited.
Due to this limitation, we only consider the high energy LHC phase for
this channel.
The process we are
interested in is therefore
\begin{equation}
pp\to G\to U \bar{u}+\bar{U}u \to H u \bar{u}\to 2j + 2l + \cancel{E}_T.
\label{Hqq:process} 
\end{equation}
The main backgrounds are $W+\mbox{jets}$, $Z+\mbox{jets}$, $WW+\mbox{jets}$ and
$t\bar{t}+\mbox{jets}$. We have simulated them as described in
Table~\ref{backgrounds:table}. In order to have enough statistics for
the $W$+jets sample we have generated it with up to hard 2 jets, 
$p_T\geq 150$ GeV, fom matrix elements (all other jets have been
generated by the parton shower).
The cuts we propose are
\begin{itemize}
\item At least 2 and no more than 6 jets.
\item Exactly 2 charged leptons, both with $p_T(l)\geq 50$ GeV and
  $|\Delta \phi(l_1,l_2)|\leq 0.5$.
\item A veto on b-tagged jets (no jet should be tagged as a b-jet).
\item $p_T(j_1)>400$ GeV, $p_T(j_2)>200$ GeV ($j_{1,2}$ denote the
  two hardest jets).
\item A cut on the invariant mass of the two charged leptons $15\mbox{
  GeV}\leq m_{ll} \leq 70$ GeV.
\item A cut on the transverse mass of the Higgs decay products
  $m_T(l,l,\cancel{E}_T) < 120$ GeV. Where the transverse mass is
  defined as
\begin{equation}
m_T=\sqrt{(E_T^{ll}+\cancel{E}_T)^2-|\mathbf{p}_T^{ll}+\cancel{\mathbf{p}}_T|^2},
\end{equation}
with $E_T^{ll}=\sqrt{|\mathbf{p}^{ll}_T|^2+m_{ll}^2}$,
$|\cancel{\mathbf{p}}_T|=\cancel{E}_T$, and $|\mathbf{p}_T^{ll}|=p_T^{ll}$. 
\item The following cut on $S_T$ using $n_j$ the number of jets between 2 and 6
  from the first cut and $n_l=2$ as a function of the test $M_G$
\begin{equation}
S_T>1.5,~2.1~,2.3\mbox{ TeV for }M_G=2,~2.5,\geq 3\mbox{ TeV}.
\end{equation}
\end{itemize}
\begin{table}[ht]
\begin{center}
\begin{tabular}{|c|c|c|c|}
\hline
Cut & $\epsilon_{M_G=2~\mathrm{TeV}}$ & $\epsilon_{W}$& $\epsilon_{t\bar{t}}$
\\
\hline
$2\leq n_j\leq 6$ & 99 & 76 & 96 \\
$n_l=2,~p_T(l)\geq 50$ GeV, $|\Delta \phi|\leq 0.5$ & 28 & 0.0206 & 0.158 \\
$n_b=0$ & 95 & 86 & 22 \\
$p_T(j_1)>400$ GeV, $p_T(j_2)>200$ GeV & 79 & 31 & 4.8 \\
$15 < m_{ll} < 70$ GeV & 91 & 47 & 75 \\
$m_T (\mathrm{Higgs})<120$ GeV & 62 & 19 & 17  \\
$S_T$ & 97 & 98 & 97 \\
\hline 
Total & 12 & $3.6\times 10^{-4}$ & $2\times 10^{-4}$
\\ \hline
\end{tabular}
\caption{Cut by cut efficiencies for the signal (benchmark model with
  MFV and $M_G=2$ TeV) and the main backgrounds
  ($W+$ jets and $t\bar{t}$) in the $Hjj$ channel 
at $\sqrt{s}=14$ TeV. All efficiencies are reported as per cent.
\label{efficiencies:Hjj}  }
\end{center}
\end{table}
The effect of the cuts on the different backgrounds and our signal for
the benchmark model, MFV realization, with $M_G=2$ TeV are described
in Table~\ref{efficiencies:Hjj}. The cuts
completely kill the $Z$ and $WW$ backgrounds which are
therefore not reported. The cut on $S_T$ has no effect for this mass
but is relevant for heavier masses. 
The corresponding statistical significance, taking into account the $BR(H\rightarrow W W^{*} \rightarrow l\nu l \nu)\approx 0.015$,
 is 
\begin{equation}
\mathcal{S}(81,85)=7.8\quad
(\mathcal{L}=100~\mathrm{fb}^{-1},~\sqrt{s}=14~\mathrm{TeV},~M_G=2\mbox{ TeV, MFV}).
\end{equation}
The global efficiencies for the signal and the
main backgrounds as a function of the test mass are given in
Table~\ref{efficiencies:Hjj:global}.
\begin{table}[ht]
\begin{center}
\begin{tabular}{|c|c|c|c|c|}
\hline
$M_G$ & $\epsilon_{M_G=2~\mathrm{TeV}}$ & $\epsilon_{W}$&
$\epsilon_{t\bar{t}}$ & $\epsilon_{WW}$
\\
\hline
1.5 & 10.4 & 0.00148 & 0.00096 & 0.01
\\ \hline
2 & 11.75 & 0.000361 & 0.0002 & 0
\\ \hline
2.5 & 6.82 & $7.4\times 10^{-5}$ & $1.57 \times 10^{-5}$ & 0
\\ \hline
3 & 7.26 & $4.54 \times 10^{-5}$ & 0 & 0
\\ \hline
3.5 & 8.15 & $4.54 \times 10^{-5}$ & 0 & 0 
\\ \hline
\end{tabular}
\caption{Global efficiencies for the signal and main backgrounds as a
  function of the test mass in the $Hjj$ channel 
at $\sqrt{s}=14$ TeV. All efficiencies are reported as per cent.
\label{efficiencies:Hjj:global}  }
\end{center}
\end{table}

\section{Results and Discussion\label{results}}

The details of our analyses and the results for the benchmark model
have been discussed in full detail in the previous section. We now
proceed to report our results as a function of the most relevant input
parameters. We have found that the discovery limits or exclusion
bounds are not very sensitive to the composite Yukawa couplings
$Y_\ast$. The main sensitivity is to the composite coupling of the
heavy gluons, $g_{\ast\,3}$, and the degree of compositeness of the
$u_R$, parameterized by $s_u$. In the next three subsections we
describe our results as a function of these parameters in the MFV
scenario for the $Ht\bar{t}$ and $H jj$ channels and in the 
anarchic scenario for the $Ht\bar{t}$ channel, respectively.
In all cases we show our results in the form of 
contour plots of the required luminosity for a
$5\sigma$ discovery, defined as $\mathcal{S}(s,b)=5$, see
Eq.~(\ref{cl}). 
We also show
contours of the luminosity required for the expected $95\%$ exclusion
bound. We have computed this bound by requiring
\begin{equation}
\mathrm{CL}_s\equiv \frac{\mathrm{CL}_{s+b}}{\mathrm{CL}_{b}}\leq 0.05,
\qquad
\bigg[
\mathrm{CL}_x=P(n\leq n_{\mathrm{obs}}|x)
\bigg],
\end{equation}
with $P(n,x)$ the Poisson distribution and in order to set the
expected bound we have fixed the observed number of events to $b$ (the
number of background events).
Finally, we also report current exclusion limits from other searches
as discussed in Section~\ref{constraints}.

\begin{figure}[ht]
\begin{center}
\includegraphics[width=0.45\textwidth]{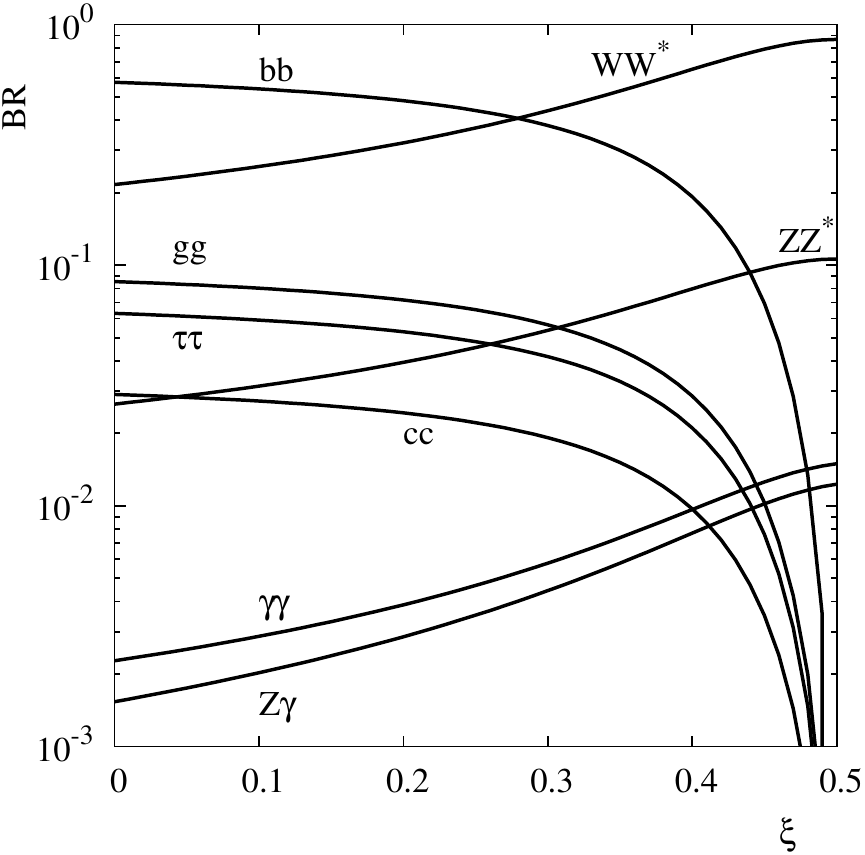} 
\hfil
\includegraphics[width=0.46\textwidth]{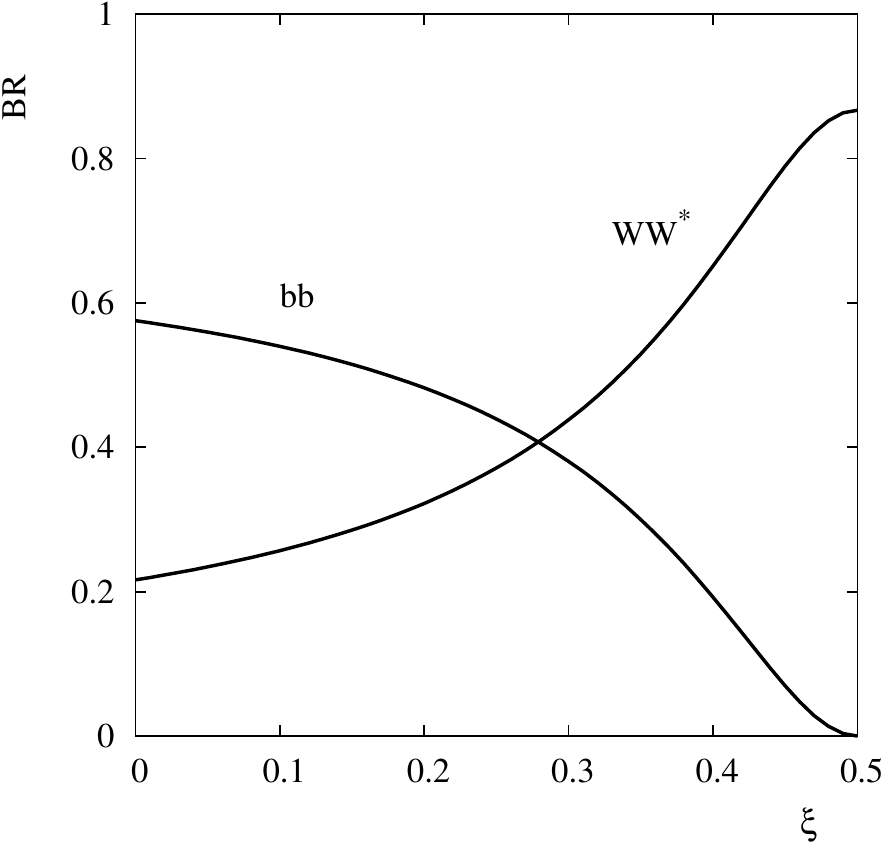} 
\caption{Branching ratios of a composite Higgs of mass 125 GeV as
  a function of the degree of compositeness in logarithmic, left
  panel, and linear, right panel, scales (for the latter only the two
  channels used here are shown).
}
\label{HiggsBRvsxi:fig}
\end{center}
\end{figure}
These results can be easily translated to different values of the
degree of Higgs compositeness ($\xi$). The number of background events
for a fixed luminosity and center of mass energy can be obtained from
the cross sections in Table~\ref{backgrounds:table} 
and the efficiencies in Tables~\ref{efficiencies:Htt:lowE}-\ref{efficiencies:Hjj}. 
The number of signal events can then be inferred from the value of
$\mathcal{S}(s,b)$ or $\mathrm{CL}_s$ in
the figures in the next three subsections. This number of signal events can then
be re-scaled by the ratio of Higgs branching ratio in the
corresponding channel for the different values of $\xi$ and the new
discovery reach or bound can be computed. To make this scaling easier
we display in Fig.~\ref{HiggsBRvsxi:fig} the different Higgs branching
ratios as a function of $\xi$.

\subsection{$Ht\bar{t}$ channel: MFV scenario}

\begin{figure}[ht]
\begin{center}
$
\begin{array}{cc}
\includegraphics[width=0.41\textwidth]{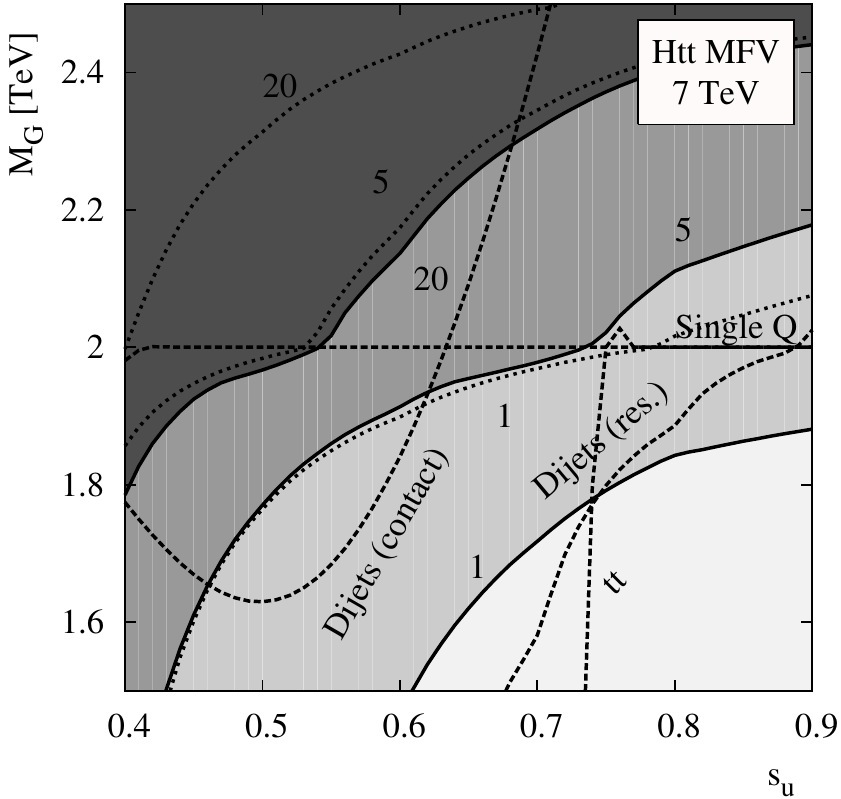} 
&
\includegraphics[width=0.41\textwidth]{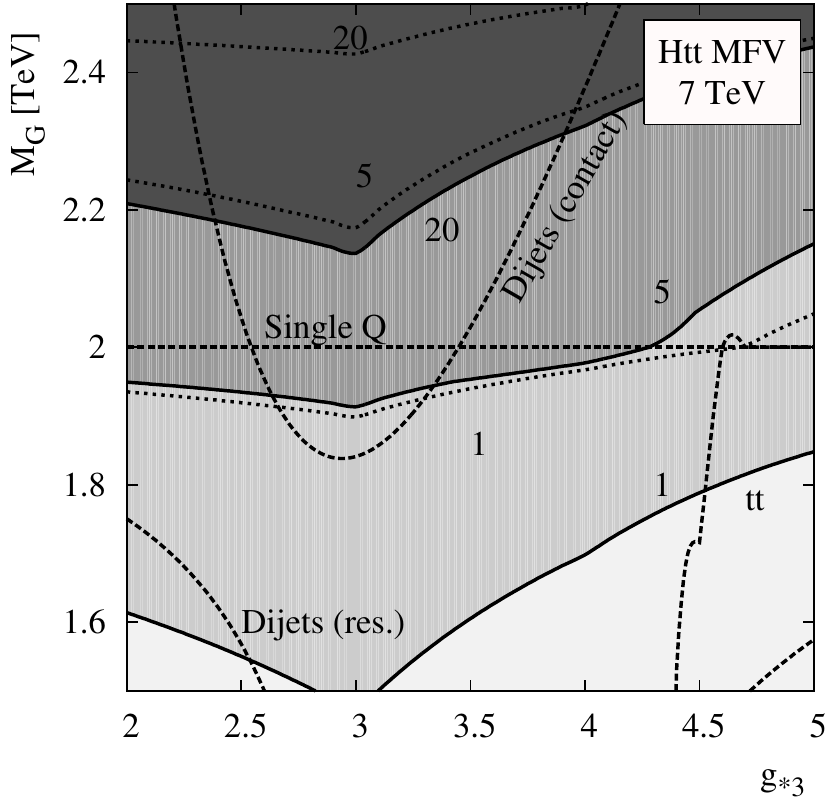} 
\\
\includegraphics[width=0.41\textwidth]{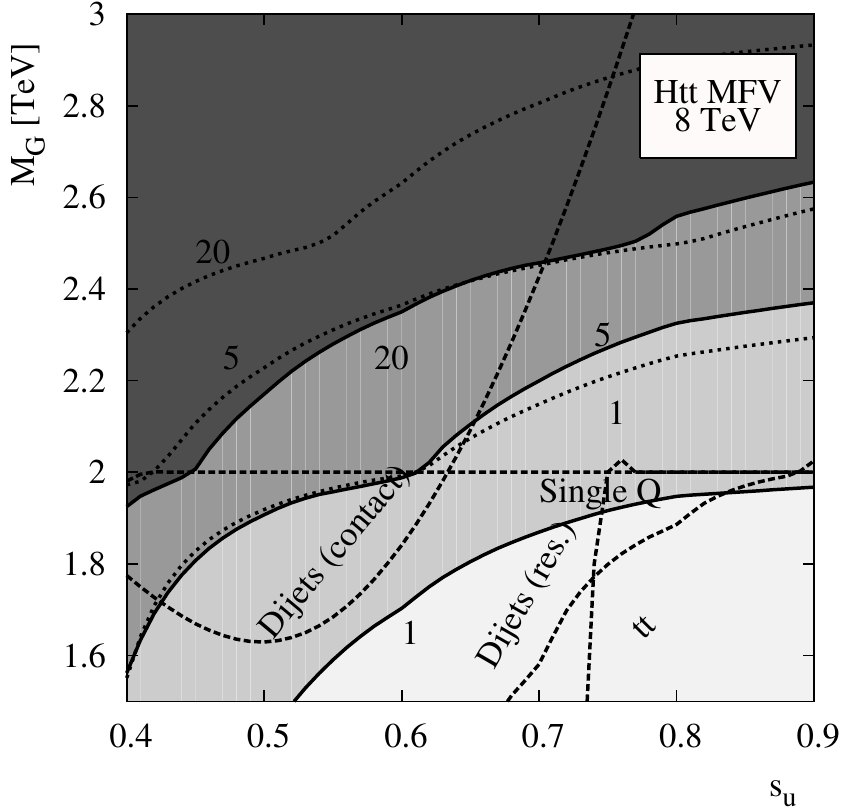} 
&
\includegraphics[width=0.41\textwidth]{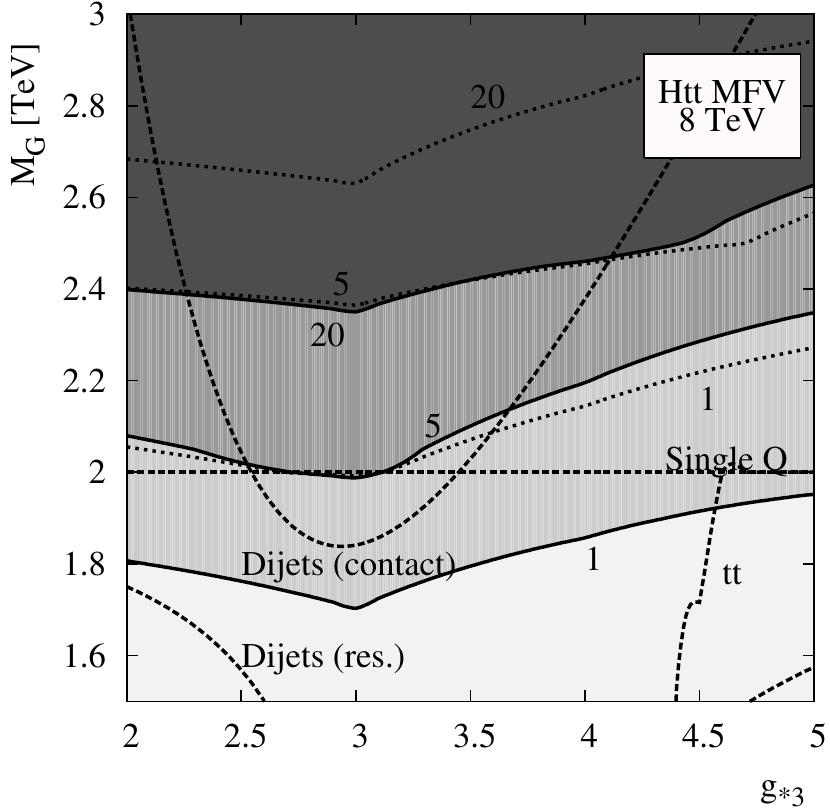} 
\\
\includegraphics[width=0.41\textwidth]{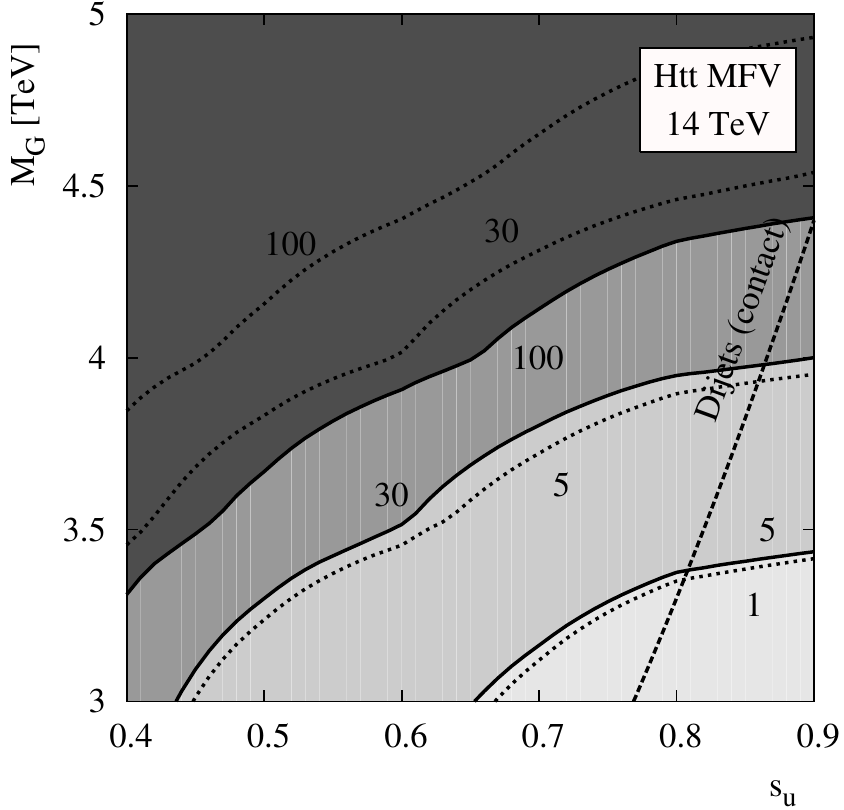} 
&
\includegraphics[width=0.41\textwidth]{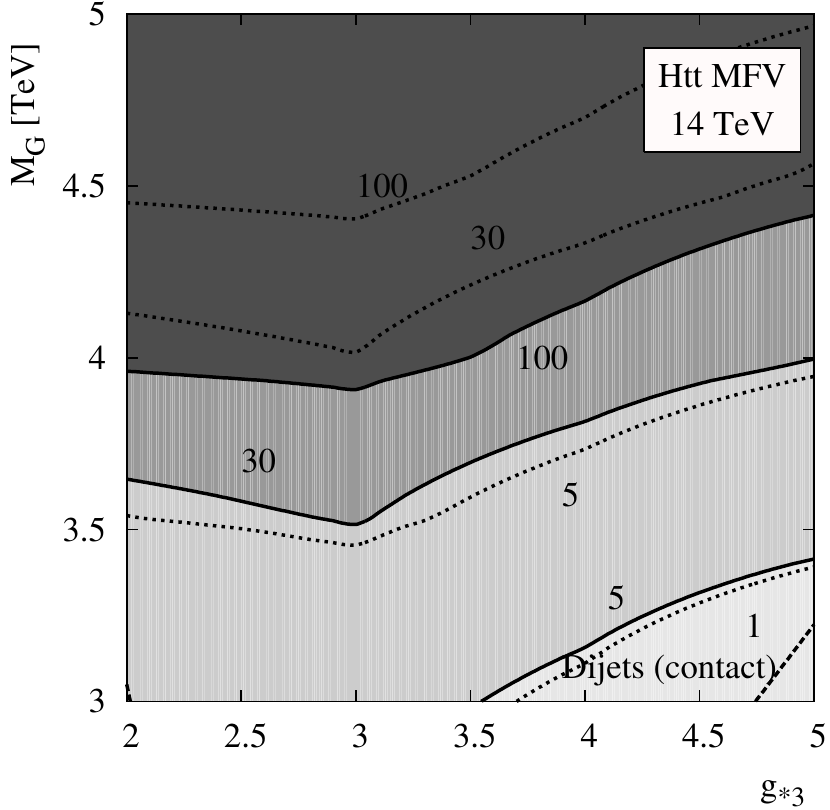} 
\end{array}
$
\caption{Contours of required luminosity for a $5\sigma$ discovery
  (bands and solid lines) and $95\%$ exclusion limits (dotted lines)
  as a function of $s_u$
  and $M_G$ (left column) and $g_ {\ast\,3}$ and $M_G$ (right column)
  for $\sqrt{s}=7,~8$ and $14$ TeV (first, second and third rows,
  respectively) in the $Ht\bar{t}$ channel (MFV scenario). 
Current bounds are shown with dashed lines (the area
  below the dashed lines is excluded). 
}
\label{results:htt:MFV:fig}
\end{center}
\end{figure}

Our main results for the $Ht\bar{t}$ channel in the MFV scenario are shown in
Fig.~\ref{results:htt:MFV:fig}, 
as a function of $s_u$ and $M_G$ (left column) 
and as a function of $g_{\ast\,3}$ and $M_G$ (right
column). The three rows of plots correspond, from top to bottom, to
$\sqrt{s}=7,~8$ and $14$ TeV, respectively. 
The discovery (exclusion) contours are represented by solid (dotted) 
lines. Current constraints are represented with dashed lines.
The excluded region is the area below the
corresponding line. Single production searches~\cite{Aad:2011yn} are
quite restrictive, independently of the input parameters. Dijet
limits on quark contact interactions (we have found that direct dijet
resonance searches lead to weaker constraints) 
can be also very restrictive,
particularly for large values of the $u_R$ degree of compositeness and
for very small or very large values of $g_{\ast\,3}$. The summary of
our results for this channel is the following:
\begin{itemize}
\item Current constraints on the model would not allow for a 5$\sigma$
  discovery with the 2011 data set at $\sqrt{s}=7$ TeV. However, 
  $95\%$ exclusion bounds could be set in the region $M_G\sim 2-2.3$
  TeV, $s_u \sim 0.5-0.7$ and $g_{\ast\,3}\sim 2.5-4$ within the
  currently allowed region of parameter space.
\item Considering now the 2012 $\sqrt{s}=8$ TeV run, a much larger
  region of the parameter space currently allowed by experimental data
  can be explored. For example $M_G\sim 2.5$ TeV can be discovered
  (excluded) with an integrated luminosity of 20 fb$^{-1}$ (5
  fb$^{-1}$). 
\item Things become even more interesting at $\sqrt{s}=14$ TeV. The region
  of masses $M_G\sim 3-4.5$ TeV can be discovered with an integrated
  luminosity of $\sim 5-100$ fb$^{-1}$. Exclusion bounds in the case that no
  signal is discovered can go up to $M_G\sim 3.4,~4,~4.5$
  and 5 TeV for integrated luminosities of $\sim 1,~5,~30$ and $100$
  fb$^{-1}$, respectively. 
\end{itemize}

\subsection{$Hjj$ channel: MFV scenario}

\begin{figure}[!ht]
\begin{center}
$
\begin{array}{cc}
\includegraphics[width=0.42\textwidth]{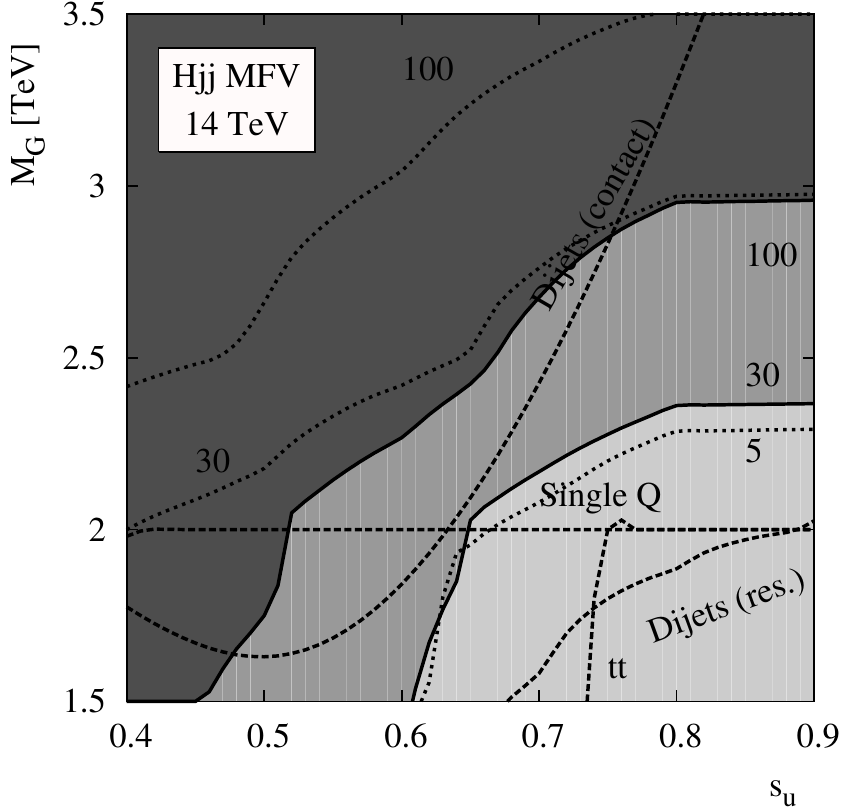} 
&
\includegraphics[width=0.42\textwidth]{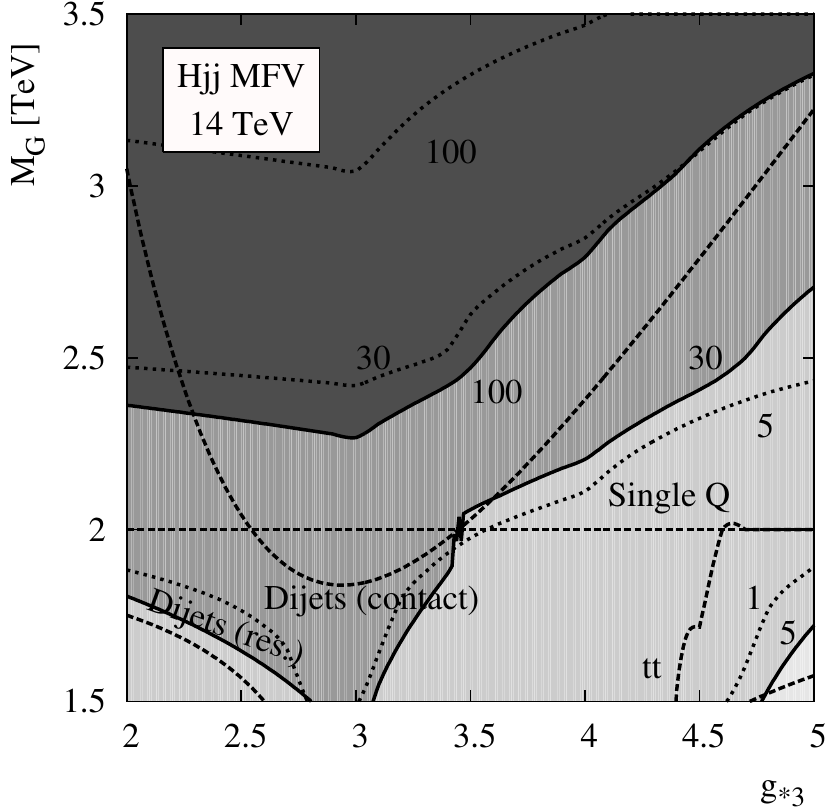} 
\end{array}
$
\caption{Contours of required luminosity for a $5\sigma$ discovery
  (bands and solid lines) and $95\%$ exclusion limits (dotted lines)
  as a function of $s_u$
  and $M_G$ (left column) and $g_ {\ast\,3}$ and $M_G$ (right column)
  for $\sqrt{s}=14$ TeV in the $H jj$ channel (MFV scenario). 
  Current bounds are shown with dashed lines (the area
  below the dashed lines is excluded). 
}
\label{results:hjj:fig}
\end{center}
\end{figure}
The discovery and 95 $\%$ bound contours for the $Hjj$ channel as a
function of $(s_u,M_G)$ and $(g_{\ast\,3},M_G)$ in the MFV scenario
are given in 
Fig.~\ref{results:hjj:fig}. 
Because of the reduced statistics due to
the leptonic decays of both $W$, this channel is not as promising as
the $H t\bar{t}$ one. In fact, even with $\sqrt{s}=14$ TeV, 
more than 30 fb$^{-1}$ of integrated
luminosity are required for discovery in the allowed region of
parameter space. With 100 fb$^{-1}$, masses up to $M_G\sim 3.3$ TeV
can be discovered and up to $M_G\sim 3.5$ TeV excluded if no signal of
new physics is observed. Thus, although this channel remains an
important complementary test of the MFV scenario, it is likely that a
much earlier signal of new physics would appear in other observables,
like dijet or single vector-like production searches.

\subsection{$Ht\bar{t}$ channel: anarchy scenario}

\begin{figure}[!ht]
\begin{center}
$
\begin{array}{cc}
\includegraphics[width=0.42\textwidth]{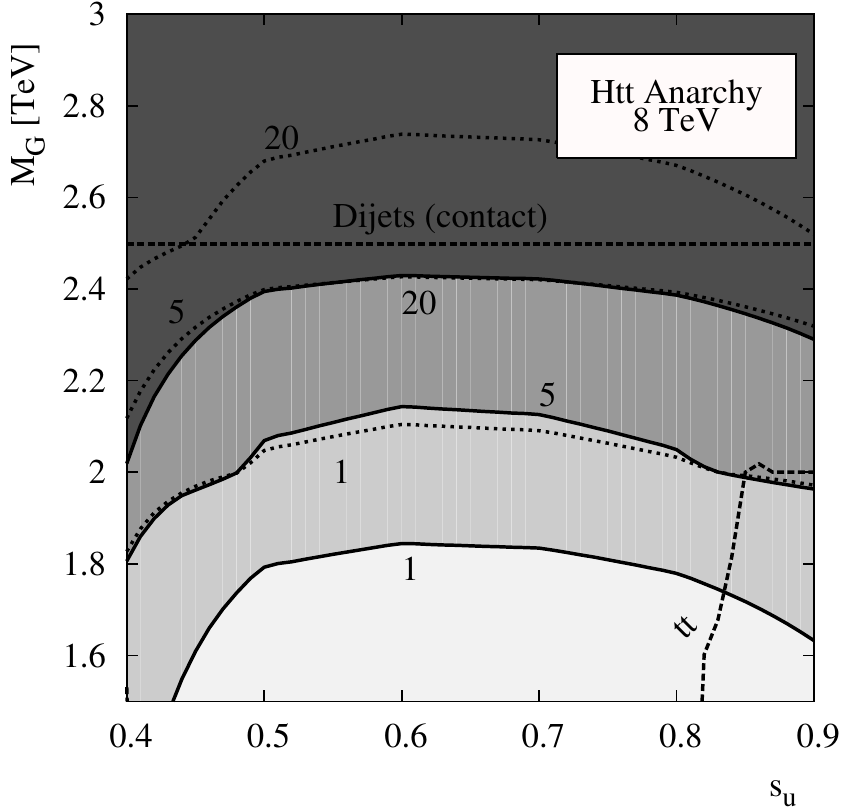} 
&
\includegraphics[width=0.42\textwidth]{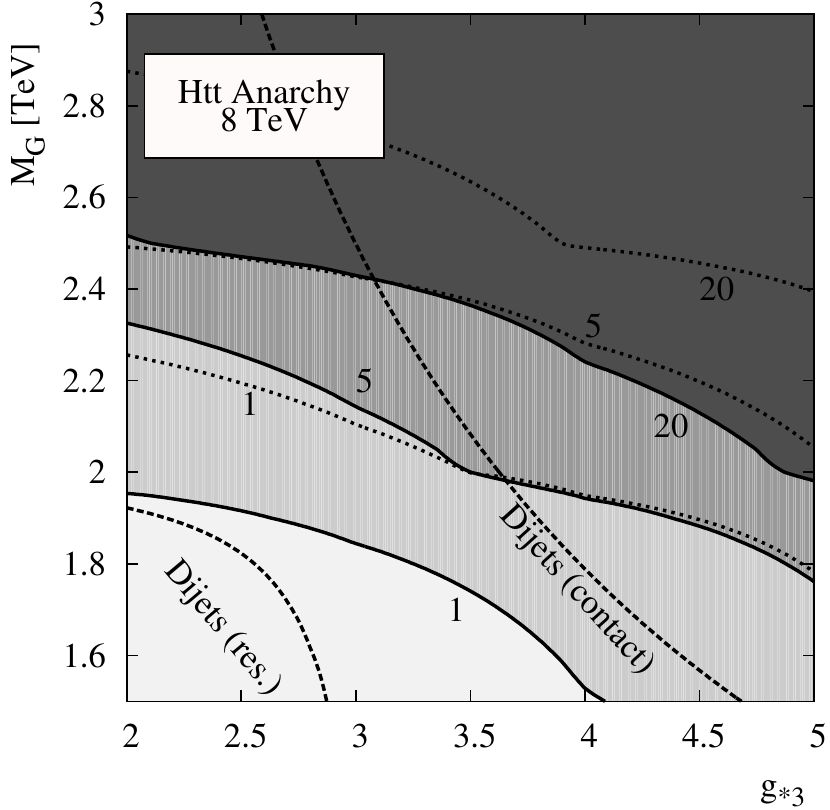} 
\\
\includegraphics[width=0.42\textwidth]{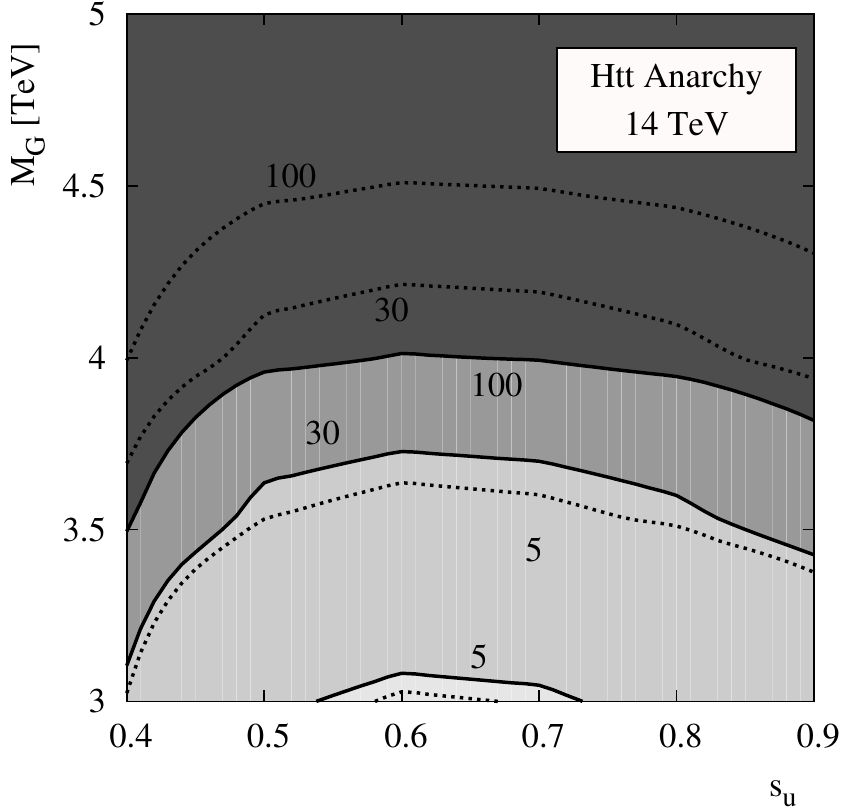} 
&
\includegraphics[width=0.42\textwidth]{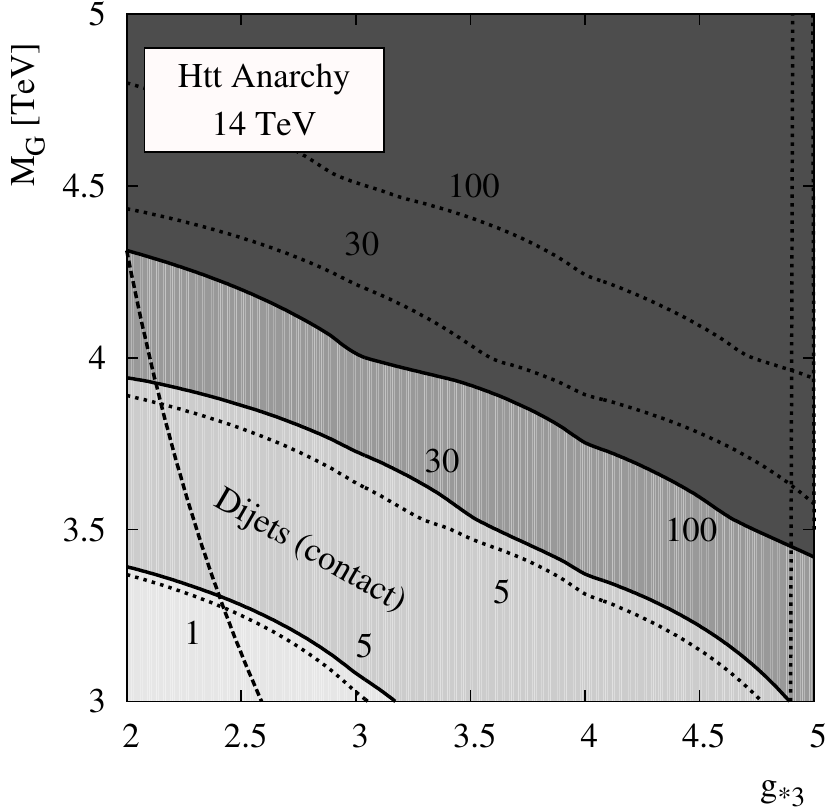} 
\end{array}
$
\caption{Contours of required luminosity for a $5\sigma$ discovery
  (bands and solid lines) and $95\%$ exclusion limits (dotted lines)
  as a function of $s_u$
  and $M_G$ (left column) and $g_ {\ast\,3}$ and $M_G$ (right column)
  for $\sqrt{s}=8$ and $14$ TeV (first and second row,
  respectively) in the $Ht\bar{t}$ channel (anarchy scenario). 
Current bounds are shown with dashed lines (the area
  below the dashed lines is excluded). 
}
\label{results:htt:Anarchy:fig}
\end{center}
\end{figure}
In the anarchic scenario, the light SM quarks are essentially
elementary ($s_u\ll 1$ for the first two generations). This has two
main implications. First, the bounds from electroweak single
production are irrelevant (the corresponding value of $\tilde{\kappa}$
is negligibly small). Second, the bounds from dijet contact
interactions depend only on $g_{\ast\, 3}$. In particular, for the
benchmark value $g_{\ast\, 3}=3$ they imply a constant bound $M_G\geq
2.5$ TeV. This bound decreases as $g_{\ast\,3}$ increases. For
instance it becomes $M_G\geq 1.5$ TeV for $g_{\ast\,3}\approx 4.6$.
Our main results are shown in Fig.~\ref{results:htt:Anarchy:fig} and
can be summarized in the following points:
\begin{itemize}
\item Using the 2011 run, masses up to $M_G\approx
  1.9-1.6$ TeV can be discovered in the region allowed by current
  constraints for $g_{\ast\,3}\gtrsim 4-5$. Exclusion
  bounds in the $M_G\sim 2.2-1.9$ TeV can be reached for
  $g_{\ast\,3}\sim 3-5$. These results assume $s_u\sim 0.5-0.7$
  (notice that 
  in the anarchy case this refers to the $t_R$ degree of
  compositeness), outside this range, the reach decreases as shown in
  the left column of Fig.~\ref{results:htt:Anarchy:fig}. 
  The plot corresponding to this energy is not shown as it is
  quite similar to the one at $\sqrt{s}=8$ TeV, only with the numbers
  reduced to match the results we have described.
\item The expected 2012 run with 20 fb$^{-1}$ at $\sqrt{s}=8$ TeV can
  lead to a discovery in the region $M_G\sim 2.5-2$ TeV (and a
  similar exclusion with just 5 fb$^{-1}$) for $g_{\ast\,3}\sim
  3-5$. Exclusion bounds in the $M_G\sim 2.8-2.4$ TeV region can be
  set, for $g_{\ast\,3}\sim 2.5-5$, with the same luminosity.
\item Data with $\sqrt{s}=14$ TeV can probe a much larger region of
  parameter space. Values up to $M_G\sim 4.3$ TeV can be discovered with
  100 fb$^{-1}$ and bounds up to $4.8$ TeV can be set with the same
  luminosity.  
\end{itemize}
It is interesting to point out the differences between the anarchic
and the MFV scenarios. In the former, discussed in this sub-section,
the reach improves for intermediate values of $s_u$ (this is just due
to the dependence of the $G t T$ and $G t \tilde{T}$ couplings) and
for small values of $g_{\ast\,3}$ (due to the larger coupling to
valence quarks and therefore larger production cross section in that
case). In the MFV case, on the other hand, the reach improves for
larger values of $s_u$ (due to the larger coupling of $G$ to valence
quarks and therefore to a larger production cross section) and for
larger values of $g_{\ast\,3}$ (due to a larger coupling of $G$ to
$u_R$ and therefore to a larger production cross section). 
Thus, for instance in
the benchmark models, the reach is slightly better in the anarchy
model than in the MFV one (although the constraints from dijets are
also stronger). The reach is even larger for smaller values of
$g_{\ast\,3}$ whereas it worsens quite a bit with respect to the MFV
scenario for large values for $s_u$ or $g_{\ast\,3}$.

\section{Conclusions \label{conclusions}}

The discovery of the Higgs boson and the measurement of its main
properties have become the major goal of the ATLAS and CMS experiments for the
2012 run.
Already the data collected during 2011 is starting to constrain many
models of physics beyond the SM. We have shown that in models with a
strong EWSB sector, the presence of new resonances of the composite
sector can mediate new Higgs production mechanisms. In particular,
single production of new vector-like quark resonances mediated by
color octet vector resonances produce an $H t\bar{t}$ or $H jj$
final state that can be easily discovered at the LHC.
Although these same final states are already present in the SM, their
kinematical features and the couplings they depend on are completely different. 
For instance, the $Ht\bar{t}$ channel is not directly
related to the top Yukawa coupling, as happens in the SM
contribution. Similarly, the two jets in the $H jj$ channel are quite
hard and central, as opposed to vector-boson fusion Higgs
production in the SM.

The experimental study of these new production mechanisms is important
for two reasons. First, it uses the LHC community Higgs effort
to explore ingredients of new models that go beyond the Higgs sector
itself. Second, it shows that a re-analysis of channels already
present in the SM but with fresh point of view can in some cases
represent main discovery channels for physics beyond the SM.

We have found that masses for new color octet
vector resonances up to 2.8 TeV can be probed with the 2011 and 2012
data sets. This enters the region currently preferred by electroweak
precision constraints. With the energy upgrade to $\sqrt{s}=14$ TeV,
up to $M_G\sim 5$ TeV can be probed with 100 fb$^{-1}$. This reach is
comparable or even better than the one of more traditional
searches~\cite{Agashe:2006hk,Lillie:2007yh}.
For masses above 2.5 TeV, boosted techniques have proven to be very
efficient in extracting the signal. We have used a very simple
analysis based on the invariant mass of fat jets but there is clearly
room for improvement with the use of more sophisticated tools.

\acknowledgments It is a pleasure to thank  F. del Aguila, 
O. Domenech, C. Grojean, J. M. Lizana, P. Masjuan and M. P\'erez-Victoria for useful discussions. This work has been supported
by MICINN projects FPA2006-05294 and FPA2010-17915, through the FPU
programme and by Junta de Andaluc\'{\i}a projects FQM 101, FQM 03048
and FQM 6552.

\bibliographystyle{JHEP}
\bibliography{myrefs}{}

\end{document}